 \documentclass[12pt, preprint]{aastex}

\newcommand{\euve}{\textit{EUVE}}

\slugcomment{The Astrophysical Journal, in press}

\shorttitle{Are Coronae Heated by Flares? II.}
\shortauthors{G\"udel et al.}

\begin{document}

\title{Are Coronae of Magnetically Active Stars Heated by Flares? \\
    II. EUV and X-Ray Flare Statistics and the Differential Emission 
        Measure Distribution}

\author{Manuel G\"udel and Marc Audard\altaffilmark{1}}
\affil{Paul Scherrer Institut, W\"urenlingen and Villigen, CH-5232 Villigen PSI, Switzerland}
\email{guedel@astro.phys.ethz.ch, audard@astro.columbia.edu }

\author{Vinay L. Kashyap and Jeremy J. Drake}
\affil{Harvard-Smithsonian Center for  Astrophysics, 60 Garden Street, Cambridge, MA 02138, USA}
\email{vkashyap@cfa.harvard.edu, jdrake@cfa.hardvard.edu}

\and

\author{Edward F. Guinan}
\affil{Department of Astronomy and Astrophysics, Villanova University, Villanova, PA 19085, USA}
\email{edward.guinan@villanova.edu}

\altaffiltext{1}{Present address: Columbia Astrophysics Laboratory, Columbia University, 
       550 West 120th Street, New York,  NY 10027, USA}
       
\begin{abstract}
We investigate the EUV and X-ray flare rate distribution in radiated energy of the late-type 
active star AD Leo. Occurrence rates of {\it solar} flares 
have previously been  found to be distributed in energy according to a power law, 
$dN/dE \propto E^{-\alpha}$, with a power-law index $\alpha$ in the range 1.5$-$2.6. 
If $\alpha \ge 2$, then
an extrapolation of the flare distribution to low flare energies may be sufficient to heat
the complete observable X-ray/EUV corona. 

We have obtained long  observations
of AD Leo with the {\it EUVE} and {\it BeppoSAX} satellites.
Numerous flares have been detected, ranging over almost two orders of magnitude in their
radiated energy. We compare the observed light curves 
with light curves synthesized from model flares that are distributed
in energy according to a power law with selectable index $\alpha$. Two methods are
applied, the first comparing flux distributions of the binned data, and the
second using the distributions of photon arrival time differences in the unbinned 
data (for {\it EUVE}).  Subsets of the light curves are tested individually, and
the quiescent flux has optionally been treated as a superposition of flares
from the same flare distribution. We find acceptable $\alpha$ values between 2.0$-$2.5 for
the {\it EUVE} DS and the {\it BeppoSAX} LECS data. Some variation is found depending on whether or not
a strong and long-lasting flare occurring in the {\it EUVE} data is included.
The {\it BeppoSAX} MECS data indicate a somewhat shallower energy distribution (smaller $\alpha$) 
than the simultaneously observed LECS data, which is attributed to the harder 
range of sensitivity of the MECS detector and the increasing peak temperatures
of flares with increasing total (radiative) energy.  The results suggest that flares can
play an important role in the energy release of this active corona. We discuss
caveats related to time variability, total energy, and multiple power-law distributions.
Studying the limiting case of a corona that is entirely heated by a population
of flares, we derive an expression for the time-averaged coronal differential emission measure
distribution (DEM) that can be used as a diagnostic for the flare energy distribution. The
shape of the analytical DEM agrees with previously published DEMs from observations of
active stars.
\end{abstract}


\section{Introduction}

The physics of coronal heating remains one of the most fundamental
problems in stellar (and solar) astrophysics. The subject has been reviewed extensively
from the point of view of theoretical concepts \citep{ionson85, narain90, zirker93},
observational solar physics  (e.g., \citealt{benz94}), and stellar physics 
(e.g., \citealt{haisch96}),
where the cited work stands exemplary for a large body of literature 
available. It is somewhat surprising that the nature of the ``coronal heating mechanism(s)'' 
still eludes agreement given high-resolution imaging of 
solar coronal structures or large statistical samples of stellar coronal
X-ray observations. For example, there is no unequivocal agreement on whether 
all, or any, of the X-ray coronal energy detected from {\it certain} classes 
of  stars is magnetic in origin. 

Coronal heating is of particular interest to stellar astrophysics since
it  relates directly to our understanding of coronal structure and 
dynamics, information that is usually obtained by means of indirect modeling.
Apart from heating models involving acoustic heating, for example on F-type stars
(see \citealt{mullan94}; although the resulting X-ray flux would be much smaller 
than observed - see \citealt{stepien89}), the 
currently advocated mechanisms are of two types: (i) Steady heating mechanisms,
e.g., by steady electric current dissipation or MHD waves, and (ii)  heating
by explosive energy release, e.g., coronal flares. The latter are attractive
heating agents since  flares  do heat plasma efficiently,
although only episodically  since the radiative and conductive losses rapidly
cool plasma to pre-flare levels, typically within minutes to hours. 

The flare heating hypothesis has gained momentum in particular from solar,
but also from stellar observations during recent years. If the quasi-steady
(``quiescent'') coronal emission is to be explained by flare contributions, flares
must act as {\it stochastic} heating agents.  \citet{parker88} proposed
that shuffling of magnetic field
footpoints in the photosphere by the convective motions leads to
tangled  magnetic field lines in the corona and thus to current sheets. With increasing
winding of magnetic fields, the necessary energy may be transported
into the coronal magnetic field where it is released by sudden relaxation 
involving reconnection.  Parker estimates that
energy dissipation occurs in packets involving $10^{24} - 10^{25}$~ergs
(``nanoflares'').  The flare-heating hypothesis 
resolves to the basic question of whether or not the {\it statistical
ensemble} of flares (in time and energy)  suffices to heat 
the apparently nonflaring coronae.

\section{Statistical Flare Observations: A Brief Overview}

Solar observations have provided evidence  for the presence of
numerous  small-scale flare events occurring in the solar corona 
at any time \citep*{lin84, porter95, gary97, krucker98}. Their distribution
in energy was found to be a power law \citep*{datlowe74, lin84} of the form
\begin{equation}\label{e:powerlaw}
\frac{dN}{dE} = k E^{-\alpha} 
\end{equation}
where $dN$ is the number of flares (per unit time) with a total 
energy (thermal or radiated) in the  interval [$E,E+dE$]. The power-law index $\alpha$ is
crucial: If $\alpha\ge 2$, then the energy integration (for a given 
time interval),
\begin{equation}
E_{\rm tot} = \int_{E_{\rm min}}^{\rm E_{\rm max}}{dN\over dE}EdE  \approx 
    {k\over \alpha-2}E_{\rm min}^{-(\alpha - 2)}
\end{equation}    
(assuming $E_{\rm min} \ll E_{\rm max}$ and $\alpha > 2$ for the last 
approximation) diverges for $E_{\rm min} 
\rightarrow 0$, i.e., by extrapolating the power law 
to sufficiently small flare energies (microflares with $\sim 10^{27}-10^{30}$~erg, 
nanoflares with $\sim 10^{24}-10^{27}$~erg), {\it any} energy release
power can be attained \citep{hudson91}. This is not the case for $\alpha <2$.
Evidently, then, one needs to measure the energy distribution of 
a statistically relevant number of flares. Solar studies have repeatedly
resulted in $\alpha$ values on the order of 1.6--1.8 for ordinary
solar flares \citep*{crosby93},
insufficient for a coronal microflare-heating model. On the other
hand, such values have proven interesting for statistical modeling since
they are also found from avalanche models \citep{lu91}.  However, 
 the implicit  assumption is that the same power law continues to
very low energies with unchanged $\alpha$. If $\alpha$ steepens for low
energies, then the microflare-heating concept may well be in order \citep{hudson91},
and such a steepening is supported by simulations of avalanche models for small events
\citep{vlahos95}.
Detailed studies of small events in the soft X-ray (SXR) and extreme
ultraviolet (EUV) ranges with 
Yohkoh, SoHO, and TRACE still provide a somewhat ambiguous picture 
\citep{shimizu95,krucker98,aschwanden00,parnell00}, 
although recent statistical investigations
suggest $\alpha = 2.0 - 2.6$ for small flares in the quiet solar corona 
\citep{krucker98,parnell00}. An extrapolation to unseen 
microevents  is always required to explain  all of the coronal energy release, 
in some cases by several orders   of magnitude. \citet{aschwanden00}
argued that the extrapolation is limited if small flares follow some scaling laws established
at higher energies, i.e., their temperatures drop to sub-coronal values, their
densities become unreasonably small, and their surface area filling factor approaches unity
possibly above the low-energy limit required.

There are two critical questions to ask here: (i) Is the power-law index of the flare
energy distribution different for different flare energy ranges? Even if 
$\alpha < 2$ for ordinary flares, a steepening at lower energies to $\alpha \ge 2$ 
can have important consequences for the coronal heating energy budget, and
the recent findings for microflares seem to support this hypothesis. (ii) 
Does the derived power-law index $\alpha$ depend on the spectral range
in which the observations were made? Incidentally, the recent microflare
observations that result in $\alpha \ge  2$ were mostly  obtained in the EUV
range where the dominant emission comes from hot plasma. It is by no means
evident that the hard X-ray production is proportional to the thermal flare
energy \citep*{feldman97}, and it is clear that the  energy released
in the hard X-ray (HXR) range (that has often been used for flare statistics) 
is a very small fraction of the total flare energy. 
Heating and particle acceleration systematics in the flare energy release may
affect both problem areas: The flare temperature and
the production of non-thermal particles systematically depend on the total
flare energy (e.g., \citealt*{feldman95}; \citealt{porter95,aschwanden99, krucker00}). 
We will address both questions in the course of the present study. 
A summary of a few relevant observations of solar flare energy distributions 
is given in  Table~\ref{flarestatistics}. A further important problem visible
in this table relates to peak flux measurements: These are reliable 
indicators for the total flare energy only if the flare duration
does not depend on the total flare energy \citep{crosby93,porter95}.

\begin{deluxetable}{lllll} 
\tablecaption{PREVIOUS MEASUREMENTS OF FLARE ENERGY DISTRIBUTIONS (SELECTION) \label{flarestatistics} }
\tabletypesize{\scriptsize}
\tablewidth{0pt}
\tablehead{
\colhead{Measurement} & \colhead{Photon energy range}  & \colhead{Flare energy}& \colhead{$\alpha$} & \colhead{Reference}    } 
\startdata
{\it Solar:}          &                         &                       &                    &                       \\
Thermal energy        & EUV \ion{Fe}{9},\ion{}{10},\ion{}{12} & $10^{25}-3\times 10^{26}$~erg & 2.3--2.6   & Krucker \& Benz 1998  \\
Thermal energy        & EUV \ion{Fe}{9},\ion{}{10},\ion{}{12} & $3\times 10^{23}-10^{26}$~erg & 2.0--2.6   & Parnell \& Jupp 2000  \\
Thermal energy        & EUV \ion{Fe}{9},\ion{}{10},\ion{}{12} & $10^{24}-2\times 10^{26}$~erg & $1.79\pm 0.08$ & Aschwanden et al. 2000   \\
Thermal and radiated energy & SXR                & $10^{27}-10^{29}$~erg & 1.5--1.6           & Shimizu 1995       \\
Peak flux             & SXR                     & $10^{25}-10^{29}$~erg & 1.4--1.8           & Shimizu 1995       \\
Peak flux             & 1--6~keV                & normal flares         & 1.75               & Drake 1971            \\
Radiated energy       & 1--6~keV                & normal flares         & 1.44               & Drake 1971            \\
Peak flux             & 1.5--12~keV             & GOES A2$-$C8 flares   & 1.88$\pm 0.21$     & Feldman et al. 1997  \\
Peak flux             & 3.5--5.5~keV \& UV      & microflares           & 2.18--2.23         & Porter et al. 1995  \\
Peak flux             & 7.7--12.5~keV           & normal flares         & 1.84\tablenotemark{a} & Hudson et al. 1969\tablenotemark{a}    \\
Peak flux             & 10--300~keV             & normal flares         & 1.8                & Datlowe et al. 1974   \\
Peak flux             & 13--600~keV             & HXR microflares       & $\sim 2.0$         & Lin et al. 1984       \\
Peak count rate       & 25--500~keV             & normal flares         & 1.8                & Dennis 1985           \\
Peak count rate       & 26--500~keV             & normal flares         & 1.54               & Aschwanden \& Dennis 1992\tablenotemark{a} \\
Peak count rate       & $>$25~keV               & normal flares         & 1.73$\pm 0.01$     & Crosby et al. 1993   \\
Peak flux             & $>$25~keV               & normal flares         & 1.59$\pm 0.01$     & Crosby et al. 1993    \\            
Peak count rate       & $>$25~keV               & normal flares         & 1.7--1.9           & Bai 1993 \\
Peak flux             & $>$30~keV               & normal flares         & 1.8--2.2           & Bromund et al. 1995  \\            
\tableline
{\it Stellar:}        &                         &                       &                    &                       \\
Radiated energy, M dwarfs&  0.05--2~keV         & $10^{30.6}-10^{33.2}$~erg & 1.52$\pm 0.08$ & Collura et al. 1988     \\
Radiated energy, M dwarfs&  0.05--2~keV         & $10^{30.5}-10^{34.0}$~erg & 1.7$\pm 0.1$   & Pallavicini et al. 1990 \\
Radiated energy, RS CVn  & EUV                  & $10^{32.9}-10^{34.6}$~erg & 1.6	     & Osten \& Brown 1999\\
Radiated energy, two G dwarfs & EUV             & $10^{33.5}-10^{34.8}$~erg & 2.0--2.2       & Audard et al. 1999\\
Radiated energy, F-M dwarfs & EUV               & $10^{30.6}-10^{35.0}$~erg & 1.8--2.3       & Audard et al. 2000\\
Radiated energy, 3 M dwarfs  & EUV              & $10^{29.0}-10^{33.7}$~erg & 2.2--2.7       & Kashyap et al. 2002\\
Radiated energy, AD Leo & EUV \& 0.1--10~keV    & $10^{31.1}-10^{33.7}$~erg & 2.0--2.5       & present work \\
\enddata
\tablenotetext{a}{quoted in Crosby et al. 1993}
\end{deluxetable}

There has been some interesting progress also on the stellar side. Most of the evidence
for an important role of flares relates to {\it magnetically active} stars with
coronal energy loss rates orders of magnitude larger than  the Sun's. 
\citet{doyle85} and \citet{skum85}
noted that the quiescent coronal X-ray luminosity of active stars, $L_{\rm X}$,
is correlated with their time-averaged U-band flare luminosity, which may indicate
that the statistical ensemble of flares is responsible for the
energy deposition in the corona. Analogously, \citet{audard00} find that the X-ray
flare frequency (above a given energy limit) also correlates with $L_{\rm X}$.
\citet{wood96}  interpret broadened ultraviolet-line wings as being due to a 
large number of transient, explosive events similar to those previously 
identified in the solar corona  \citep*{dere89}. 
Observations of active stars  often show slow variability  
within a factor of 2 over both short and long terms, apart from  obvious flaring 
\citep*{ambrust87, collura88, gudel96, kashyap99, sciortino99}. Statistically,
the hotter X-ray emitting plasma component is more variable than the cooler one, 
indicating that flares may be involved \citep{giam96}. 
Further, small flare events with energies of the order
of $10^{27} - 10^{28}$~ergs have become observable with the Hubble Space
Telescope in cool M dwarfs, and some of their energy distributions suggest $\alpha > 2$ 
\citep{robinson95, robinson99, robinson01} although we note that these transition region
events may not be related to the physics of coronal heating. 

Studies of {\it statistical flare energy distributions} in stellar coronae have been rare, due 
to the paucity of relevant data sets (Table~\ref{flarestatistics}). \citet{collura88} and 
\citet*{pallavicini90} derived 
$\alpha = 1.5-1.7$ for a sample of M-dwarf flare observations  with EXOSAT, and
\citet{osten99} report  $\alpha = 1.6$ for a set of flares on RS CVn binary systems observed
with {\it EUVE}. All observations obtained flare energies 
simply by integrating count rates during flares selected from the light curves. 
Also, these investigations collected flare information from
numerous different stars, regardless of distance (introducing selection bias for small flares),
luminosity (and hence overall magnetic activity level), binarity, etc.
To avoid possible selection bias,
\citet*{audard99} and \citet{audard00} applied a flare search algorithm to {\it EUVE} light
curves of individual active main-sequence stars, taking into account flare superpositions and
various binning to recognize weak flares. Although the statistical confidence 
intervals are broad, the results indicate a predominance of relatively steep power laws
including $\alpha \ge 2$, suggesting that the extension of the 
flare distribution to lower energies may contribute considerably
to the detected energy loss. 

In statistical flare studies,
the identification of weak flares close to the apparently quiescent emission 
level becomes an ill-defined problem. 
Explicit detection methods discriminate against small flares due to overlap with
larger flares, confusion between the many approximately simultaneous small flares, or
short detection times above the significance level \citep{hudson69,audard00}.
Limited signal-to-noise ratios add to the problem.  Some of these complications can be overcome 
to some extent by fully modeling the superposition of a statistical ensemble of flares and
comparing observable quantities between models and observations. An important diagnostic of
this kind may be the coronal differential emission measure distribution (DEM) since
it is determined by the ensemble of plasma packets heated to different temperatures.
A time-integrated observation of a time-dependent heating mechanism may therefore reveal
a characteristic DEM that contains diagnostics for the stochastic flare distribution 
(\citealt{gudel97, gudeleal97}; the modeling described in the latter papers concentrated
on small $\alpha$ with important effects due to large and long-lasting flares). 
We will derive an analytical expression for  a DEM of a flare-heated corona in this paper.
The present work and its companion paper \citep{kashyap02} are, on the other hand,
predominantly concerned with statistical light curve analysis, comparing model
light curves with long monitoring observations obtained in the EUV and X-ray ranges. 
While \citet{kashyap02} focus on a detailed instrumental modeling of the {\it EUVE} DS 
instrument and a statistical analysis of the photon arrival time distribution for a sample 
of stars, the present paper concentrates on a dedicated, long observing campaign of AD Leonis 
using the {\it EUVE} and  {\it BeppoSAX} satellites; two complementary methods, one presented
in \citet{kashyap02}, are applied, and the results are used to compute
the emission measure distribution of a continually flaring stellar corona.

The paper is organized as follows: \S 3 discusses our data reduction techniques,
and \S 4 the analysis methods. In \S 5 we present our results, and we discuss various 
features in \S 6. Finally, \S 7 presents our conclusions. Although we follow our hypothesis
that all of the observed X-ray emission is due to a statistical ensemble of flares,
it is convenient to use the term ``quiescent''  emission for the emission level at which 
individual flares can no longer be separated, i.e., for the  quasi-steady emission.

\section{DATA SELECTION AND REDUCTION}

Our target for the present investigation is the nearby dMe star AD Leo.
AD Leo is a well-studied flare star with a high flare rate. Its spectral
class is M3~V, and its X-ray luminosity amounts to log$L_{\rm X} = 28.95$~($L_{\rm X}$
in erg~s$^{-1}$). We use
a distance of $d = 4.90$~pc (see Audard et al. 2000 for further details on the target,
and for references). Its quiescent count rates in the {\it EUVE} DS, the 
{\it BeppoSAX} LECS, and the {\it BeppoSAX} MECS used here are, respectively, 
$\sim$0.15, 0.105, and 0.025 cts~s$^{-1}$ (these values refer to the mean of all flux bins
in which no  significant flares were evident.)

\begin{figure}[h!] 
\epsscale{1.0}
\plotone{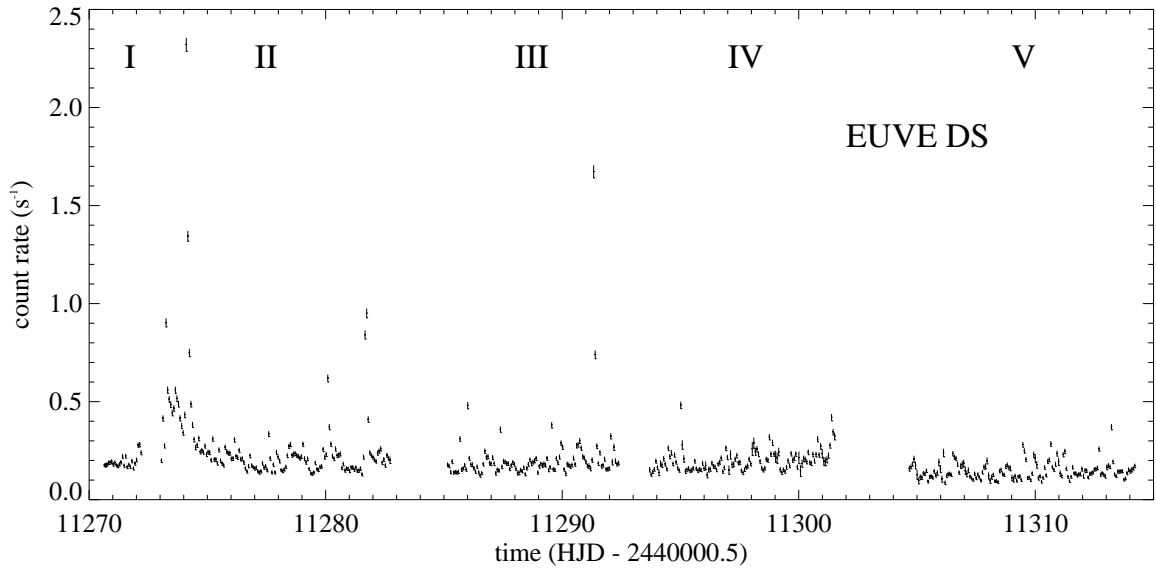}
\caption{{\it EUVE} DS light curve of AD Leo (before primbsching and dead time corrections), 
obtained between April 2, 1999, and May 16, 1999.
          Segment V suffers from ``dead spot'' reduction in effective area and from high radiation. The $1\sigma$
	  error bars
	  are typically $\pm$0.01~ct~s$^{-1}$ and have been plotted.\label{figure1}}
\end{figure}

The \textit{Extreme Ultraviolet Explorer} \citep[\euve, e.g.][]{malina91} 
data presented here were obtained in several segments between 1999 April 2 and 1999 May 15 
(see Table~\ref{obslog}).
DS Remapped Archive QPOE files were rebuilt 
using the \texttt{euv1.8} package within IRAF. 
For our method 1 (see below) in which we used binned data,
we applied primbsching (telemetry saturation) and deadtime corrections to all data sets, and excluded
time intervals for which the combined correction factors exceeded 30\%.
To avoid fluctuations that may still occur within
one satellite orbit window owing to residual inaccuracies in these
corrections (e.g., due to the South
Atlantic Anomaly), we decided to generally
bin  data  to one bin per orbit (5663~s), which turns out to be sufficient to 
recognize numerous EUV flares, to resolve them in time, and to provide
a good signal-to-noise ratio per bin. The light
curve is shown in Figure~\ref{figure1}. The last time interval
(see Table~\ref{obslog}) suffered from too high radiation, with correction
factors higher than typically acceptable. This segment was {\it not} used in 
the subsequent analysis (for any of the methods).  The final data
set used for method 1 contained 470 bins. 

\begin{deluxetable}{llccc}
\tablecaption{OBSERVING LOG (1999 APRIL/MAY)\label{obslog}}
\tabletypesize{\scriptsize}
\tablewidth{0pt}
\tablehead{
\colhead{Segment}  &\colhead{Instrument} & \colhead{from UT}         & \colhead{to UT}  & \colhead{HJD range} \\
\colhead{}  & \colhead{}          & \colhead{MM/DD  hh:mm} & \colhead{MM/DD  hh:mm}  & \colhead{-2440000.5}  } 
\startdata
I   & {\it EUVE}             & 04/02 16:09 & 04/04 04:54 & 11270.673--11272.204 \\
II  & {\it EUVE}             & 04/05 00:46 & 04/14 16:41 & 11273.032--11282.695 \\
III & {\it EUVE}             & 04/17 03:32 & 04/24 09:27 & 11285.147--11292.394\\
IV  & {\it EUVE}             & 04/25 16:29 & 05/04 11:41 & 11293.687--11302.487 \\
V   & {\it EUVE}             & 05/06 16:41 & 05/16 05:11 & 11304.695--11314.216 \\
\tableline
I   & {\it BeppoSAX}         & 05/01 06:34 & 05/03 04:48 & 11299.274--11301.200 \\
II  & {\it BeppoSAX}         & 05/08 05:47 & 05/10 13:10 & 11306.241--11308.549 \\
III & {\it BeppoSAX}         & 05/12 08:39 & 05/15 15:05 & 11310.360--11313.628 \\
\enddata
\end{deluxetable}

\begin{figure}
\plotone{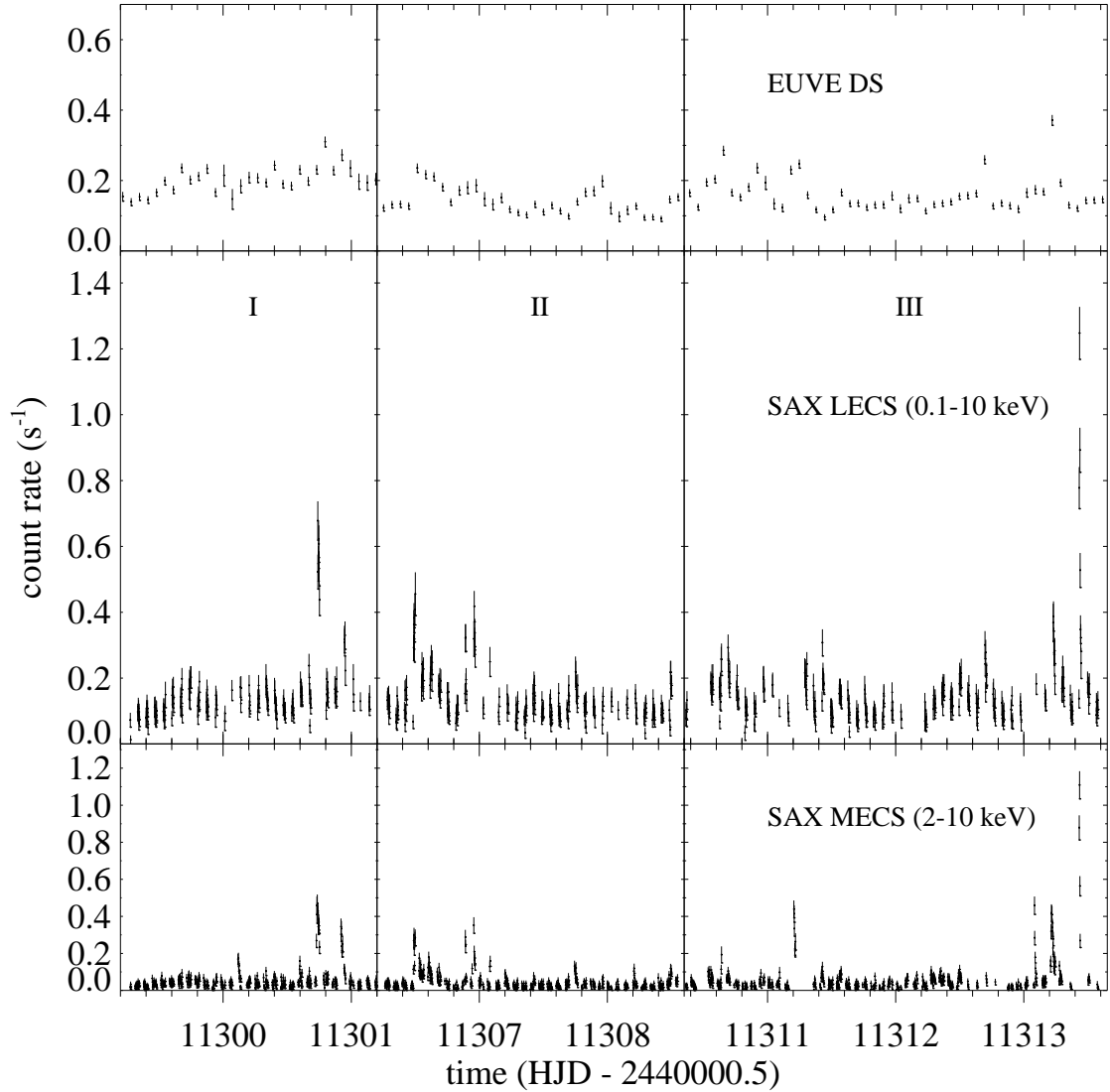}
\caption{{\it BeppoSAX} MECS and LECS light curves, compared with simultaneous
     {\it EUVE} DS data. Roman numerals refer to the {\it BeppoSAX} segments 
     (see Table~\ref{obslog}). Note that {\it EUVE} DS data obtained during {\it BeppoSAX}
     intervals II and III are unreliable due to high radiation background and were not 
     used.  
     \label{figure2}}
\end{figure}

The \textit{BeppoSAX} satellite \citep{boella1997a} pointed its Low and Medium Energy Concentrator 
Systems (LECS and MECS, respectively; \citealt*{parmar1997,boella1997b}) 
and its Phoswich Detector System (PDS; \citealt*{frontera1997}) towards AD Leo three times 
(Table~\ref{obslog}), spanning a total time of 15~days for 270 ks of exposure time. 
Most of the observations were performed simultaneously with the \textit{EUVE} observations. No
significant signal was detected in the PDS instrument, even during flares. The three
different pointings were similar, allowing us to merge the individual data sets
into single LECS and MECS data sets. The cleaned and linearized event files from
LECS and MECS23 (MECS2 and MECS3 combined file) from the SAX Science Data Center 
pipeline processing were filtered with the good time intervals that exclude the 
events occurring while there was no attitude solution. Because of the overall faster variability
seen in the X-ray range and in order to maximize the number of available points while still
retaining a sufficient signal-to-noise ratio to detect a large number of flares, we
decided to resolve each orbit visibility interval into a few bins of length
200~s (a test performed with 1 bin per orbit provided compatible results that were, given
the small number of points, ill-constrained). We thus used 
a total of 658 bins for the LECS light curve and
1363 bins for the MECS light curve. The MECS has more bins since it observed
longer (up to about 3000~s) during each orbit, and it suffered from fewer bad 
time intervals during the on-source observations. The light curves of all three segments 
are shown in Figure~\ref{figure2}.

Average quiescent spectra were extracted for LECS and MECS, excluding obvious large flares
(but still including small flares because, in our working hypothesis, there is no
strict difference between flaring and non-flaring emission). 
The source was extracted
inside circular regions of radius  8.17$^{\prime}$ and 4$^{\prime}$ for LECS and MECS, respectively. 
Blank sky pointings were used to model the instrumental and sky X-ray backgrounds. We
checked the local background with two semi-annular regions for LECS (see 
\citealt*{parmar1999} for full details) and two regions perpendicular to 
the on-board calibration sources for MECS.
We found  that the blank-sky backgrounds are suitable for our data. 
We used the \texttt{lemat} tool in SAXDAS~2.0.1 to create LECS response 
matrices, and we adopted the standard MECS responses distributed by 
the mission (September 1997). Finally,
the data were grouped with a minimum of 25 counts per bin. Data from 0.12 to
4~keV were kept for the LECS spectrum, while data from 1.65 to 10.5~keV were
kept for the MECS23 spectrum \citep[see][]{fiore1999}. We fitted the combined
spectra in XSPEC \citep{arnaud96} using several isothermal collisionally ionized equilibrium
plasmas (the so-called MEKAL code) with a fixed interstellar hydrogen absorption 
column density $N_\mathrm{H} = 10^{18}$~cm$^{-2}$ \citep{sciortino99}. We also
introduced a free constant factor to the LECS spectrum to account for
cross-calibration discrepancies in the overall effective areas. We left the Fe abundance 
free while we kept the other abundances at solar photospheric values. A three-temperature
model provided a reasonable fit (Table~\ref{saxfit}). 
Finally, we derived an emission measure distribution (DEM) from the combined LECS and MECS 
data in SPEX, averaging the DEMs from the polynomial and the regularization methods as described in 
\citet{kaastra96a} (this is further discussed in \S~\ref{dem}).

\begin{deluxetable}{cccccccccc}
\tabletypesize{\scriptsize}
\tablecaption{FIT TO THE QUIESCENT  {\it BeppoSAX} DATA\label{saxfit}}
\tablewidth{0pt}
\tablehead{
\colhead{$\log N_\mathrm{H}$\tablenotemark{a}} & 
\colhead{$kT_1$}   & 
\colhead{$kT_2$}   &
\colhead{$kT_3$} &
\colhead{$EM_1/10^{50}$}  & 
\colhead{$EM_2/10^{50}$} & 
\colhead{$EM_3/10^{50}$} &
\colhead{Fe\tablenotemark{b}}     & 
\colhead{f\tablenotemark{c}}     & 
\colhead{$\chi^2$/dof} \\
\colhead{($\mathrm{cm^{-2}}$)} & 
\colhead{(keV)}   & 
\colhead{(keV)}   &
\colhead{(keV)} &
\colhead{($\mathrm{cm^{-3}}$)}  & 
\colhead{($\mathrm{cm^{-3}}$)} & 
\colhead{($\mathrm{cm^{-3}}$)} &
\colhead{}     & 
\colhead{}     & 
\colhead{} 
}
\startdata
$=18.0$  &$0.18 \pm 0.01$ & $0.68 \pm 0.01$ & $1.80 \pm 0.08$  & $5.4 \pm 0.4$ & $13.6 \pm 0.4$ & $4.5 \pm
0.2$  & $0.54 \pm 0.03$ & $0.69 \pm 0.03$ & 315.48/261\\
\enddata
\tablenotetext{a}{Fixed value}
\tablenotetext{b}{Abundance relative to the solar photospheric value (using values from Anders \& Grevesse
1989)}
\tablenotetext{c}{Effective-area cross-calibration factor LECS/MECS23}
\end{deluxetable}

\section{ANALYSIS}

We analyzed the calibrated and cleaned data  using two different methods. 
Since the present data quality
does not allow us to model the run of temperatures and emission measures of
individual flares (and therefore to integrate the {\it modeled} radiative
losses across complete flares), we will assume that the observed count rate
is proportional to the total X-ray flux from the star. We will
discuss the reliability of this assumption in \S 6.3. 

\subsection{Count Rate Distributions}

We  compare the count rate distributions of the light curves with simulated 
data sets composed of a
statistical flare distribution. The simulations are performed as follows:
We first calculate a statistical power-law distribution of flare energies 
(equation~\ref{e:powerlaw}) specified by three free 
parameters: the total number of flares during the simulation time (flare rate $R_f$), 
the power-law index $\alpha$, 
and a lower cut-off energy $E_0$. The latter is necessary for practical reasons but must 
also physically exist in the case of $\alpha \ge 2$. The count rate  normalization is 
arbitrary and will be re-normalized later.  The flares, initially defined as delta functions, 
are  randomly distributed in time. The length of the simulation time interval was chosen to contain
5490 bins, i.e., about ten times more than the binned {\it EUVE} and {\it BeppoSAX}/LECS light curves
and about 4 times more than the MECS light curve, i.e.,
we compare the observation with about ten (resp. four) statistical 
realizations of a  simulation of equal length. 

This model light curve is then convolved with a  calculated
exponential decay profile with pre-set decay time constant $\tau$. This latter form
is typical of that commonly observed for stellar flares.
The model flares start at their peaks (i.e., the rise phase of the flare is very short, 
compatible with the observations). 
We note that the individual flare decay times vary somewhat, also due to
superimposed flares that may  widen the apparent flare profile. Experiments
with a profile that was half as wide for the LECS and the MECS data showed, however,
that the results are quite insensitive to the precise profile shape. The flare decays
are generally considerably slower in the EUV range than in soft X-rays. Based on a detailed analysis
with fully resolved {\it EUVE} DS data sets including ours,  a decay constant
of 3000~s proves appropriate \citep{kashyap02}, and we adopt this value for consistency
with the study presented by the latter authors although we will test (as they did) 
variable decay times as well.  Most rapid features in the LECS and MECS 
required a characteristic decay time scale of only 300--400~s; we adopted a decay time of 360~s.

Since during a considerable fraction of the satellite orbit the target star is not accessible,
we also simulated the effects of ``windowing''. The DS instrument observed
AD Leo contiguously for typically 1520~s during {\it EUVE}'s 5663~s orbit, the LECS for 1400~s
and the MECS for 3000~s during the 5780~s {\it BeppoSAX} orbit. One anticipates
that correcting for this effect is  statistically of little importance.  
In particular, since the flare decays are nearly exponential, a flare from which the initial 
part is cut off behaves like a flare detected at peak time, with the same  decay time scale 
but with a smaller amplitude. In this case, flares of a given energy statistically
shift to somewhat lower energies. Since the flare decay constants are similar or equal
for all flares, their count rates are statistically suppressed by the same  factor,
resulting only in a lower normalization of the count rate distribution, which is irrelevant.
Our conjecture was verified with appropriate models that cut simulated flares arbitrarily 
the same way as the real observation does. 

\begin{figure}[ht!]
\epsscale{0.58}
\plotone{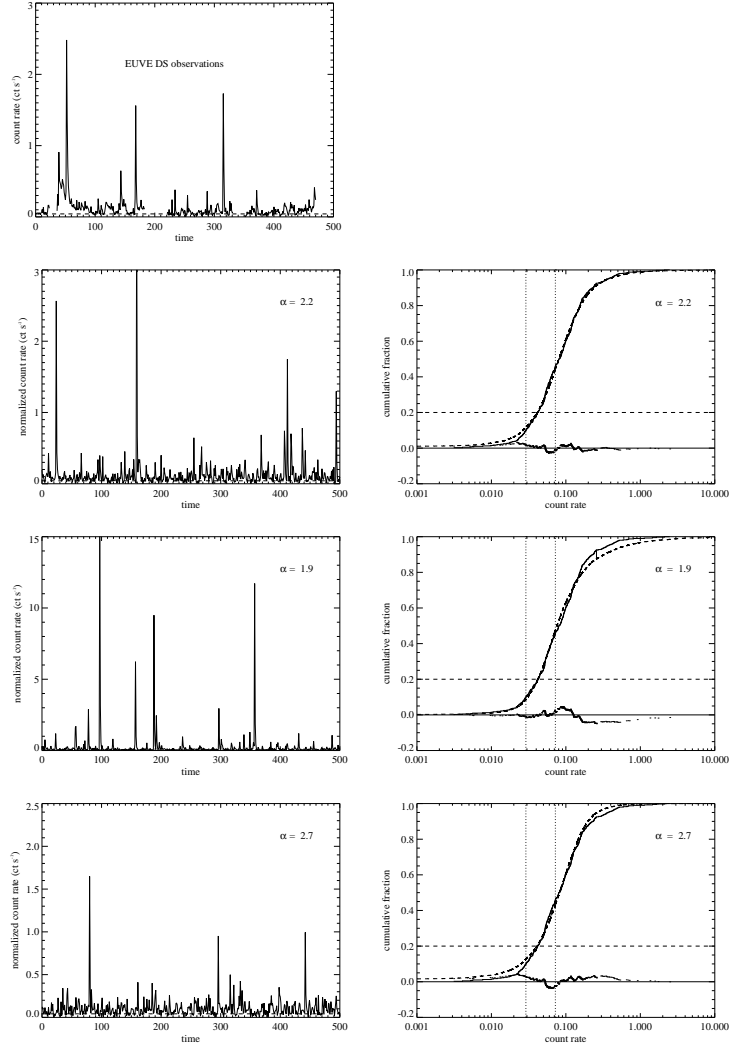}
\pagebreak
\caption{\small Examples of statistical 
          flare simulations from method 1 for different power-law
         distributions. The {\it EUVE} DS data set of AD Leo is shown. 
	 Only emission exceeding quiescent level has been modeled. The early
	 large flare is included.  Three simulated best-fit examples for different $\alpha$ are shown.
	 {\bf Top figure:} Observed light curve.
	 {\bf Second line:} Optimum case; $\alpha = 2.2$.
	 {\bf Third line:} Too shallow distribution
	  with $\alpha = 1.9$.
	 {\bf Bottom line:} Too steep distribution with $\alpha = 2.7$.
         For 2nd, 3rd, and bottom figure panel:
	 {\bf Left:} Simulated light curve (extract, normalized). Dashed line marks the
	  lower limit to count rates that were considered for optimizing the fit. 
	 {\bf Right:}
	 Cumulative count distribution for data (solid) and model (dashed), and
	 difference (dotted, around zero line). The maximum difference (vertical bar between 
	 observation and model)   is minimal for the best fit. The two vertical dotted lines
	 mark the interval for the model count rate normalization. The horizontal dashed line
	 marks the count rate level above which the differences between observation and model
	 were  considered.\label{figure3} }
\end{figure}

We then sort all  bins of the light curve, both for the observed and the model light curve, in order
of increasing count rate $c$ to obtain a cumulative count rate distribution ${\cal{N}}(<c)$
(number of bins with a count rate up to a given $c$, where we normalize
${\cal{N}}$ by the total number of bins, i.e., $0\le {\cal{N}}\le 1$; see 
Fig.~\ref{figure3}--\ref{figure6}). The cumulative distribution is 
very sensitive to systematic deviations in the shape of the count rate
distribution as produced, for example, by different flare energy distributions.
Since the observed and the model distributions are not mutually
normalized in their count rates, we renormalize the (cumulative) model 
distribution such that the average count rate within a range
$[{\cal{N}}_1,{\cal{N}}_2]$ is the same as the average count rate in  
the corresponding ${\cal{N}}$ range of the observed distribution.
At this stage,  noise corresponding to the observed Poisson noise is added to 
the model points, and the model data points are sorted again in count rate.

The low end of the count rate distribution is inherently ill-defined. It is sensitive
to statistical fluctuations from the superposition of numerous weak flares while
at the same time the relative errors are largest in that range. Further, tails
from a few long-decay flares may  modulate this count rate level in time. Also,
intrinsic non-flaring variability such as due to emerging magnetic loops or rotational modulation
can introduce slow variations. We therefore keep the final normalization interval
above zero, namely at $[{\cal{N}}_1,{\cal{N}}_2] = [0.1,0.5]$, and do not further consider the
portion at  ${\cal{N}} < 0.2\ (< 0.1$ for LECS/MECS).

For any selected $\alpha$, we generated a large family of  models that 
differ only in their flare rates. We selected the example that minimizes the largest
vertical distance between the model and the observed cumulative distributions 
(i.e., minimal max$(|{\cal{N}}_{\rm obs}-{\cal{N}}_{\rm model}|)$) as our best fit.
The traditional Kolmogorov-Smirnov test cannot be applied to the two distributions since
the normalization (one of the required fit parameters) has been derived from the properties
of the distributions themselves. We rather {\it simulate} a sample of observations with defined
properties ($\alpha$, $R_f$, and $E_0$) and a length equal to the real
observation. We then analyze each simulated observation 
precisely the same way as the real observation in order to find statistical approximations 
to the confidence limits of our results.  If the model distribution is
too shallow ($\alpha$ too small) we find an excess of large count rates, i.e., the model cumulative
distribution lies below the observed cumulative distribution toward larger count rates 
(see Fig.~\ref{figure3}--\ref{figure6} below). 
Conversely, if the model distribution is too  steep ($\alpha$ too large), then the model 
distribution lies above the observed distribution.

\subsection{Analysis of Photon Arrival Time Differences} 

\citet{kashyap02} present a detailed description of the second method that is
founded on a detailed modeling of the {\it EUVE} DS detector. 
We provide only a brief summary.  The basis of the method is
that, as count rates rise and fall with flaring activity, the
intervals between photon arrival times decrease and increase according
to the Poisson distribution appropriate for the count rate at any
given moment.  These changes in the arrival time differences ($\delta
t$) cast a signature on the photon event list that changes according
to the nature of the underlying source variability.  Thus, for a given
observation, the observed distribution $f(\delta t)$ summarizes the
character of the variability of the source during that observation.  A
fixed flare distribution (equation~\ref{e:powerlaw}) gives rise to a
definite $f(\delta t)$ provided that the observation is of sufficient
duration to contain a representative range of intensities, and
different flare distributions will give rise to different arrival-time
difference distributions.  Note that $f(\delta t)$ is not sensitive to
the actual temporal locations of the flares, but rather depends only
on the stochastic ensemble described by equation~\ref{e:powerlaw}.
The power-law index, $\alpha$, that best describes the observed light
curve is determined by comparing the observed arrival time difference
distribution $f_{obs}(\delta t)$, with simulated distributions
$f_{sim}(\delta t)$.

This method was developed to deal easily and rigorously with the
windowing inherent in low Earth orbit observations (with a typical
observing time of 30 minutes for each $\sim 90$~minute orbit), such as
those obtained by {\it EUVE}, and to allow for proper treatment of telemetry
saturation (``primbsching'') and deadtime effects  
that can introduce somewhat variable corrections during one orbit.
Since we deal here with photon lists, a proper treatment of
the corrections {\it continuously in time} is important.
The advantage of
the method is that it operates directly on the observed photon event
list, avoiding the need for time binning.  Simulated event lists can
be windowed in exactly the same way as the observed event list, and
primbsching effects can be applied to the simulated events, censoring
them in the same stochastic way as the observed events.

We assume here that the observed light curves can be described by the
sum of a flaring component, with a power-law frequency distribution of
flare energies as described by equation~\ref{e:powerlaw}, and a constant 
component. The flaring and constant components are described by the {\em flare
rate} $C_f$, and a ``quiescent background'' rate $C_b$, respectively.  As for the first
method, the flares are assumed to be impulsive events whose count rates decay
exponentially with a time scale of a few thousand seconds.  The exact 
decay time can be varied to
best match any observed flares or the observed and synthetic photon
event lists themselves.  In practice, we have found that results
obtained for decay times in the range 1000$-$5000~s are typically very
similar because the extensive overlapping of flare events tends to
decrease the sensitivity of the results to this parameter \citep{kashyap02};
the final results presented here assume a decay time of
3000~s, identical to the value found and applied by \citet{kashyap02}.
For specified values of $C_f$ and $C_b$, a synthetic light
curve corresponding to the entire interval covered by the observed
light curve can then be realized through a Monte Carlo algorithm,
assuming a random distribution of flares in time but subject to the
power-law frequency distribution of total energies.  This light curve
is then windowed by the observed photon event ``good time intervals''
and a synthetic event list is derived through a Poisson realization of
the resulting light curve.  The synthetic event list is then pruned by
discarding photons at a rate corresponding to the observed Primbsch
factor.  The remaining set is identical in its ``instrumental
characteristics'' to the observed data and the observed and synthetic
event lists can be compared directly.

Observed photon arrival time differences are compared to those
synthesized across a grid of the parameters $\alpha$, $C_f$ and $C_b$
using the $\chi^2$ statistic to compare $f_{sim}(\delta t)$ with
$f_{obs}(\delta t)$.  A number of simulations (typically $\sim 10$)
are carried out and the median value among the resulting $\chi^2$ are
used to compute the likelihood of obtaining the observed data for the
given set of parameters $\{ \alpha, C_f, C_b \}$.  In the case of AD
Leo, we have examined each of the three {\it EUVE} observation segments II, III, 
and IV independently (Table~\ref{obslog}), treating them as three different 
observations, as well as treating the whole sequence II-IV at once.  This enabled us to examine
the degree of consistency of our derivation of $\alpha$ from
segment to segment.

\section{RESULTS}
    
\subsection{Parameters and Tests}\label{tests}

Our simulated light curves were generated assuming an exponential flare decay time $\tau$
independent of the flare energy or amplitude. If the decay time
increases systematically with the total radiated  energy, e.g., $\tau \propto E^{\beta}$, 
then the flare 
amplitudes increase less than proportionally with the energies, which may
be interpreted as a steeper  energy distribution. However, increasing 
the decay time of larger flares produces more bins at large count rates (but only an equal number
of additional bins at low count rates), and this effect counteracts the apparent
steepening of the energy distribution.

Observationally, a dependence of the decay time on energy is marginal.
\citet{aschwanden00}  investigated scaling laws from solar nanoflares to larger flares,
covering 9 orders of magnitude in energy. While most of the geometric and 
physical parameters exhibit a strong scaling with the flare size, Aschwanden et al.
report that the time scale (radiative or conductive) does {\it not} depend on the flare
size. In fact, as stronger flares tend to show larger electron densities and
higher coronal peak temperatures (\S 3.1 in \citealt{aschwanden00}), they should 
radiatively or conductively decay faster than small flares, which is not observed.
The range of decay times (their Table 2) is much larger than any possible trend 
between nanoflares and large flares. \citet{feldman97} measured soft X-ray FWHM
durations of a large sample of flares over three orders of magnitude and found no
trend. The differences between flare decays are due to scatter within classes of
flares of about equal energy. \citet{shimizu95} investigated energies and durations
of small solar active-region transients and again found no clear trend in the
durations over four orders of magnitude in peak count rate. We inspected the best 
defined flares in our data although superpositions of flare
profiles may introduce ambiguity. The MECS data are best suited for such an analysis
since they provide the longest intervals of visibility per orbit, a high flare-to-quiescent 
emission contrast, and a time bin resolution of 200~s. We found no trend for a
correlation between flare amplitude and decay time; the scatter in the decay time itself
dominates. The same holds true for the other data sets. 
However, the sample of
selected large X-ray flares of \citet{pallavicini90} from {\it different} stars shows a 
weak trend if four orders of magnitude in radiated energy are included, but the total
variation is no larger 
than the scatter in duration at a given energy. For individual stars, there is no clear trend.
Their data sample can be best fitted with a relation $\tau \propto E^{0.25}$. Given that
our dynamic range (ratio between largest to smallest explicitly detected flare count rate) is of 
the order of 10$-$20 for binned data, this effect may statistically influence the results. We therefore tested
the DS, LECS, and MECS analysis by introducing variable decay times as specified above.

We have used two realizations of our analysis for the binned {\it EUVE} data (method 1)
that show a well-developed quiescent emission. First, we applied it to the data 
from which the (constant)  quiescent level was subtracted. In this case, we  test whether
we can find a model distribution 
that is compatible with the {\it observed} flare emission. Second, we applied the
method to the complete data including the quiescent emission. Since the latter, in our model, should 
be the superposition of unresolved small flares, the addition of a quiescent level simply
corresponds to the extrapolation of the power law to lower energies, and we expect that 
the power-law index does not significantly change. This second realization also allows us to derive the lower
cut-off energy $E_0$ required to explain the quiescent emission.  
The comparison between the two realizations could  potentially reveal a basic difference
between quiescent and flaring emission. If the flare energy distribution does not steadily
continue toward more numerous small flares that eventually merge with the quiescent level,
then the first realization would be subject to a cut-off energy possibly above the quiescent level
(bi-modal flux distribution), and the derived values for $\alpha$ may differ.
These two extreme cases will further be discussed below. 
The contrast between flares and the quiescent emission is much larger in the {\it BeppoSAX} data; we 
treated only the complete data sets in these cases for the primary results but also subtracted
the quiescent level for an assessment of the confidence ranges (see below).

\subsection{Basic Findings}

We first discuss results obtained using method 1 under the assumption of constant exponential decay
times $\tau$ for all flares.
Figure 3 shows results for the {\it EUVE} light curve of AD Leo (segments I$-$IV) from
our first method.  The top figure represents
the light curve binned to one point per orbit (5663~s).  Here, a constant quiescent 
count  rate of 0.13~ct~s$^{-1}$ has been
subtracted, corresponding to the lower envelope of the light curve. 
The x-axis (``time'') gives the sequential bin number, with
observing gaps at 30, 200, and 340 fully considered.  The alternative analysis in which
the initial large flare was excluded  used only bins above bin no. 70.

The three lower rows then illustrate, in this order, the best fit found in the analysis (i.e.,
optimum $\alpha$ and optimum flare rate), a selected case with too low $\alpha$ (but
again optimum flare rate), and similarly a case with too high $\alpha$. In each case, the figure
on the left shows an extract of the simulated  and normalized light curve with noise added, 
while the figure on the right shows the respective cumulative distributions. In the latter, the
solid line shows the observed cumulative count rate distribution ${\cal{N}}(<c)$ while the dashed line
illustrates the normalized count rate distribution from the model. The two dotted vertical lines indicate the
range used to normalize the model distribution  to the observation. Our statistical fit criterion
(minimizing the largest vertical distance between the two distributions) was
applied only above the dashed horizontal line (above ${\cal{N}} = 0.2$ or 0.1). The dotted function close
to and around the horizontal ${\cal{N}} = 0$ line illustrates the difference
``model -- observation''  for the cumulative distributions.  Evidently, the optimum index 
is $\alpha \approx 2.2$ in this case, while values as low as 1.9 or as high as 2.7 are considerably worse.

\begin{figure} 
\epsscale{0.58}
\plotone{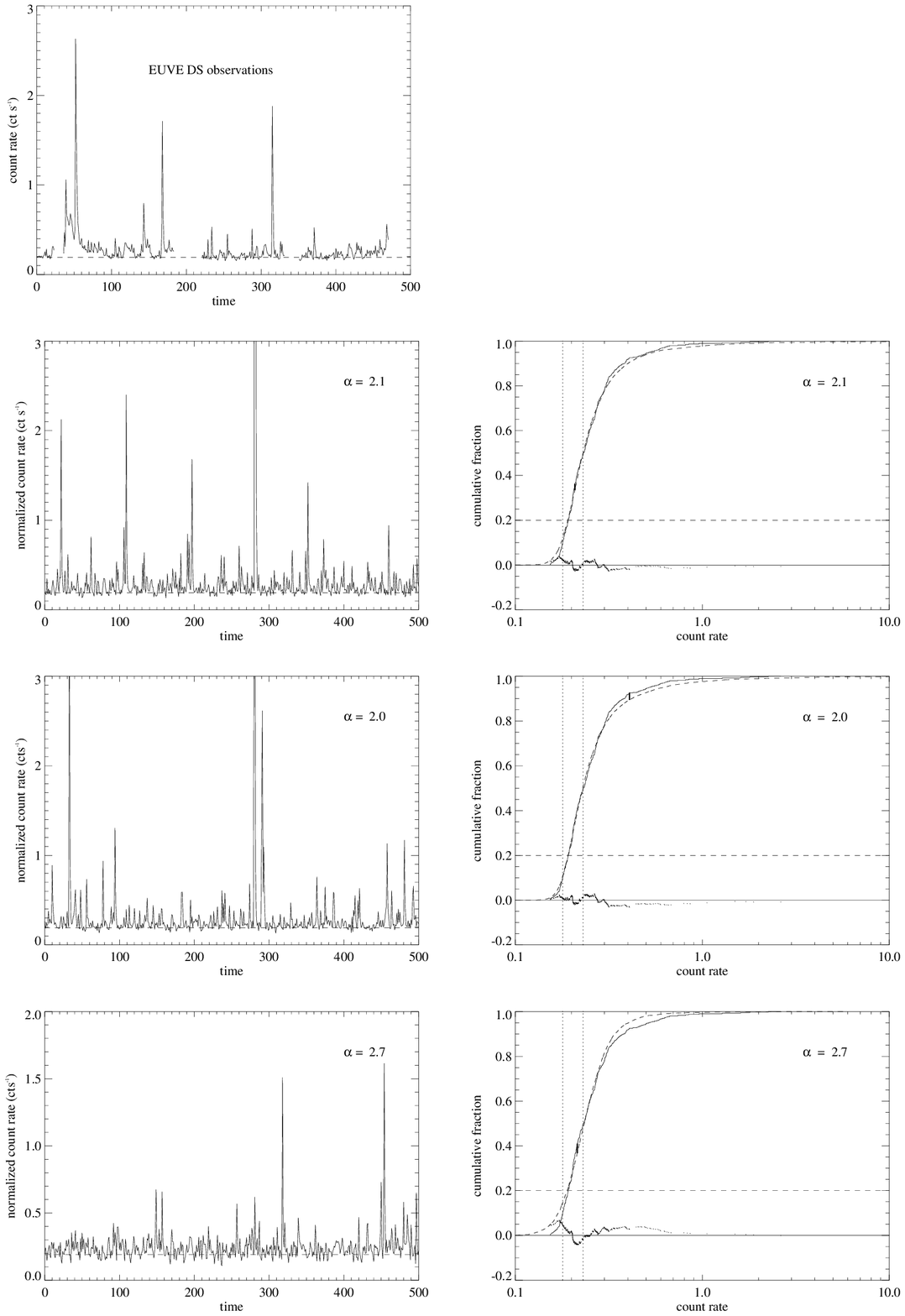}
\caption{Similar to  Fig.~\ref{figure3} (method 1), but all emission, including 
         quiescent emission,  has been modeled. 
	 {\bf Top panel:} Observed light curve.
	 {\bf Second panel:} Optimum case; $\alpha = 2.1$.
	    {\bf Third panel:} Too shallow distribution with $\alpha = 2.0$.
	  {\bf Bottom panel:} Too steep distribution with $\alpha = 2.7$.\label{figure4}}
\end{figure}

\begin{figure}
\epsscale{0.4}
\plotone{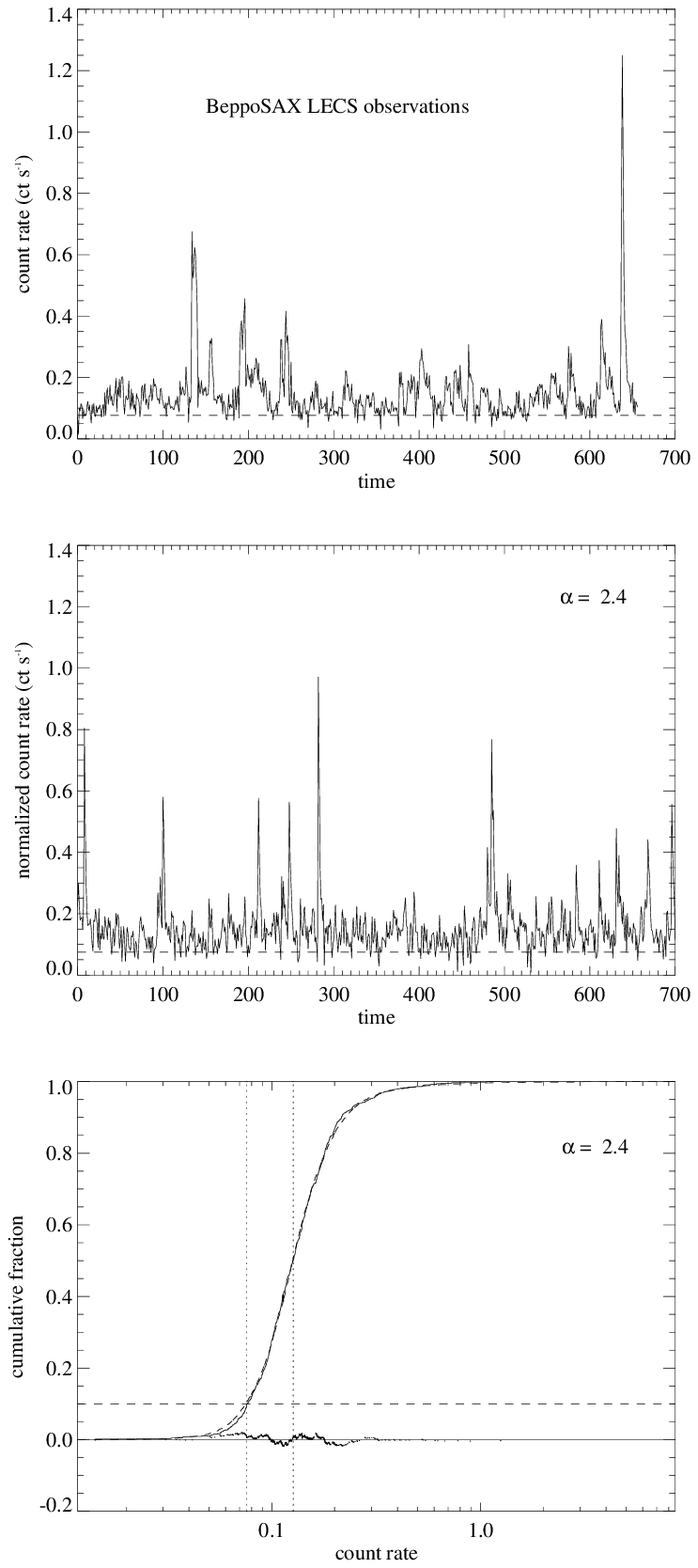}
\caption{Similar to Fig.~\ref{figure3} (method 1), for LECS. Only
        optimum model is shown, with $\alpha = 2.4$. 
	Gaps between observation segments and between orbits are not
	retained.\label{figure5}}
\end{figure}

\begin{figure} 
\epsscale{0.4}
\plotone{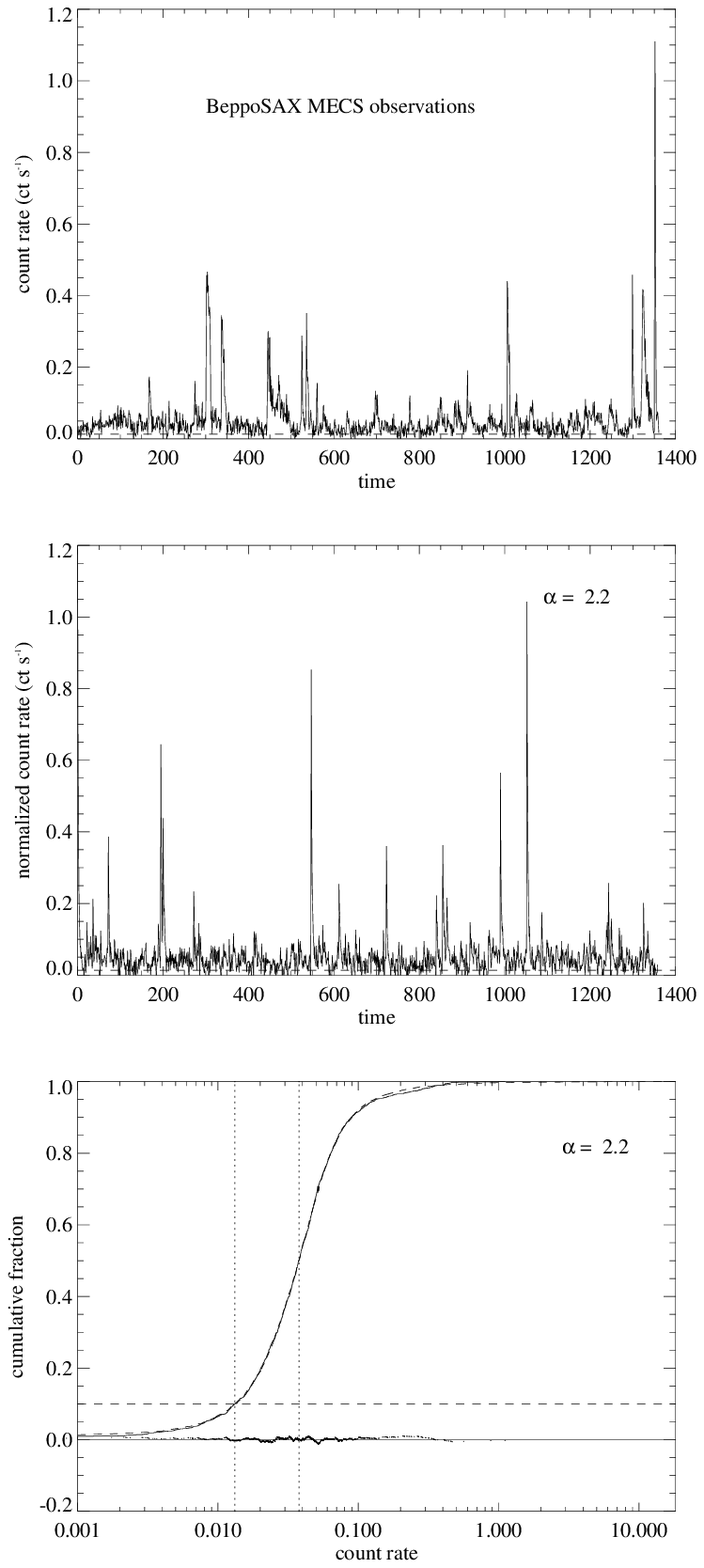}
\caption{Similar to Fig.~\ref{figure3} (method 1), for  MECS. Only
        optimum model is shown, with $\alpha =2.2$. 
	Gaps between observation segments and between orbits are not
	retained.\label{figure6}}
\end{figure}

Figure~\ref{figure4} shows the equivalent analysis but without subtraction of a quiescent level.
The optimum value  is formally found at $\alpha = 2.1$ although the cumulative count rate 
histogram suggests a somewhat higher $\alpha$. The fit is generally somewhat poor, but a test in which 
the quiescent level was reduced by 50\% still produced 
$\alpha = 2.1$.   Lastly, Figures 5 and 6 show the
results from the analysis of the LECS and MECS light curves, respectively. Only the
observed light curve, an extract of the  optimum model, and the cumulative distributions 
are shown. We find optimum values of $\alpha = 2.4$ and 2.2, respectively.

The statistical confidence ranges derived from our sample of simulated observations are as follows.
Approximately 90\% of best-fit solutions for the {\it EUVE} DS simulations (performed for $\alpha 
= 2.0$ and $\alpha =  2.2$) were found symmetrically within a range of $\pm 0.1$ centered
at the correct $\alpha$. We thus adopt $\Delta\alpha = \pm 0.1$ as our 90\% confidence limits
for the {\it EUVE} DS results. The lower signal-to-noise ratio of the {\it BeppoSAX} data
constrained the simulation results less well, probably due to the large dominance of the
noise fluctuations at low count rates. We therefore repeated the analysis (both for the real
and the simulated observations) constructing cumulative distributions from those data only that
exceed the quiescent level by $1\sigma$, thus selecting predominantly the visible 
flares only. For the LECS, $\alpha = 2.4$ was confirmed,
while for the MECS, we found optimum fits for $\alpha = 2.0$. 
From the sample of simulated observations, we again find a 90\% width of
about $\pm 0.1$, although $\alpha$ is slightly underestimated by 0.1 on average. This effect
is probably due to selectively choosing the  larger count rate bins. We conservatively
adopt a combined (statistical and systematic) uncertainty of $\pm 0.2$ for the LECS and MECS (90\%).

The following systematics are evident (Table~\ref{minenergy}): (i) In all cases, we find 
acceptable values for $\alpha$ at or above 2, with  $\alpha = (2.1-2.3) \pm 0.1$
for the DS data, $\alpha =2.4\pm 0.2$ for LECS and $\alpha = (2.0-2.2) \pm 0.2$ for MECS. 
When we excluded  the large flare in the DS light curve,
the distribution steepened by $\Delta \alpha \approx 0.1$ (Table ~\ref{minenergy}), an 
effect that is  to be  understood as follows: The presence of a large flare will 
tend to give a smaller $\alpha$ because comparatively less power will be contained in smaller
flare events. Conversely,
the selective elimination of the population of the strongest flares steepens the
distribution although a genuine single power law cannot be retained if too large a fraction
of the distribution is eliminated. The long decay  of the large flare clearly also
biases the count rate distribution toward a lower $\alpha$ (adding a large number of bins at high
count rate levels). Overall, thus, the DS and the LECS ranges for $\alpha$ agree. The MECS
range appears to be systematically lower. We will discuss
this effect  further in Section~\ref{dependence}.

\begin{deluxetable}{lrrccccc}
\tablecaption{$\alpha$ VALUES FOR AD LEO FROM METHOD 1; MINIMUM MODEL 
      FLARE ENERGIES AND LUMINOSITIES\tablenotemark{a}\label{minenergy}}
\tabletypesize{\scriptsize}
\tablewidth{0pt}
\tablehead{
\colhead{Data set,} & \colhead{bin} & \colhead{$\alpha$} & \colhead{Minimum flare}   & \colhead{Minimum flare}      & \colhead{Minimum peak}   &
\colhead{Detection limit}  & \colhead{Detection limit}           \\
\colhead{segment } & \colhead{size}  & \colhead{(90\% error)} & \colhead{counts}         & \colhead{energy}             & \colhead{luminosity}            &
\colhead{flare energy}     & \colhead{peak luminosity}           \\
\colhead{}          &\colhead{(s)}   & \colhead{}   & \colhead{}      & \colhead{(erg)}              & \colhead{(erg~s$^{-1}$)}   &
\colhead{(erg)}              & \colhead{(erg~s$^{-1}$)}
} 
\startdata
DS I--IV\tablenotemark{b} & 5663 & 2.2(0.1)  & 62     & $1.7\times 10^{31}$ & $5.8\times 10^{27}$  & $3\times   10^{31}$ &$1\times 10^{28}$ \\
DS I--IV\tablenotemark{c} & 5663 & 2.3(0.1)  & 60     & $1.7\times 10^{31}$ & $5.5\times 10^{27}$  & $3\times   10^{31}$ &$1\times 10^{28}$\\
DS I--IV\tablenotemark{d} & 5663 & 2.1(0.1)  & 0.5    & $1.4\times 10^{29}$ & $4.8\times 10^{25}$  & $3\times   10^{31}$ &$1\times 10^{28}$\\
DS I--IV\tablenotemark{e} & 5663 & 2.3(0.1)  & 3.5    & $9.6\times 10^{29}$ & $3.2\times 10^{26}$  & $3\times   10^{31}$ &$1\times 10^{28}$\\
LECS I--III               & 200  & 2.4(0.2)  & 0.8    & $3.8\times 10^{29}$ & $1.0\times 10^{27}$  & $1.4\times 10^{31}$ &$4\times 10^{28}$\\
MECS I--III               & 200  & 2.0-2.2(0.2) & 0.4 & $1.1\times 10^{30}$ & $2.9\times 10^{27}$  & $3.6\times 10^{31}$ &$1\times 10^{29}$
\normalsize
\enddata
\tablenotetext{a}{For given time bin size and for the optimum case; minimum flare counts/energy/peak luminosity refer to smallest flares used in the simulation
          with optimum KS test result. Detection limit refers to 3$\sigma$ levels for the respective bin size.}
\tablenotetext{b}{quiescent level subtracted, all data}
\tablenotetext{c}{quiescent level subtracted, large flare excluded}
\tablenotetext{d}{quiescent level included, all data included}
\tablenotetext{e}{quiescent level included, large flare excluded}
\end{deluxetable}

\begin{figure} 
\epsscale{0.3}
\plotone{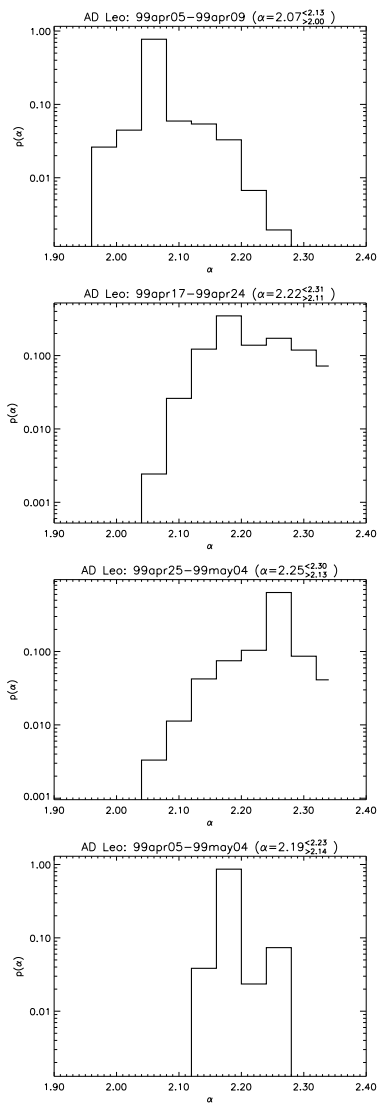}
\caption{The derived probability of the power-law index $\alpha$ matching that
of the {\it EUVE} DS observations of AD~Leo based on the 
distributions of photon arrival time differences (method 2).  The most probable
value of $\alpha$ corresponds to the peak value of each distribution
and is printed together with the corresponding 95\%\ confidence
intervals at the top of each figure. 
From top to bottom: 
{\bf a)} Segment II, including the large flare.
{\bf b)} Segment III.
{\bf c)} Segment IV.
{\bf d)} Segments II-IV combined.
\label{figure7}}
\end{figure}

The results from the second method, based on photon arrival statistics, are illustrated in
Figures~\ref{figure7}a$-$\ref{figure7}d, where the derived
probabilities of power law indices $\alpha$ matching the observed index
are plotted as a function of $\alpha$. The results are also summarized
in Table~\ref{alpha_arrival}. The first three figures
illustrate the derived probabilities for the three segments II, III, and IV treated
separately, while the fourth shows the results of the analysis of the
whole data set II-IV treated as a single observation.  The important feature
of all these figures is that the most probable value of $\alpha$ is
again always greater than 2 but less than 2.3, based on 90\%\ or 95\%\ confidence
intervals.  Segment II appears to have an 
optimum index that is slightly lower than that of the last two segments.  This is caused by
the large flare, which contains a significant fraction of the total observed
counts in that segment (see above).  The confidence intervals for the last two
segments are remarkably similar, indicating $2.1 \leq
\alpha \leq 2.3$.  The most probable value of $\alpha$ based on all
three segments  (Figure~\ref{figure7}d) is $\alpha = 2.2$. The values found here are 
thus in excellent agreement with results from the first method.

\begin{deluxetable}{lcc} 
\tablecaption{$\alpha$ VALUES FOR AD LEO FROM METHOD 2 (DS) \label{alpha_arrival} }
\tablewidth{0pt}
\tablehead{
\colhead{Segment}  &\colhead{Most probable} & \colhead{95\% Confidence} \\     
\colhead{DS}       &\colhead{$\alpha$}      & \colhead{Interval}    } 
\startdata
II         & 2.07  &          2.00--2.13   \\
III        & 2.22  &          2.11--2.31   \\
IV         & 2.25  &          2.13--2.30   \\
II--IV     & 2.19  &          2.14--2.23   \\
\enddata
\end{deluxetable}

We further investigated the case for an exponential decay time 
$\tau \propto E^{\beta} = E^{0.25}$.
In the first method, sufficiently small flares will have decay times  much smaller than
the bin length and thus the method is sensitive only to their total energy contribution,
but not to their decay time. Given our relatively coarse binning, we thus expect to find
rather stable results for $\alpha$. This is confirmed, with deviations of no
more than $\pm$0.1 in $\alpha$. The second method is much more
sensitive to details in the light curve and could, in principle, find both power-law indices
$\beta$ and $\alpha$. 
In practise, however, there is an acceptable family of solutions $(\alpha,
\beta)$. The acceptable $\alpha$ increases with increasing $\beta$, reaching values up to
2.6$-$2.8 for $\beta = 0.25$. We interpret this as being due to a larger time occupation by 
relatively large count rates, i.e., short photon arrival time differences, if the energetic 
flares are stretched in time relative
to the small flares (despite the corresponding decrease in peak count rate). The model
light curve thus appears to be ``too hard'', and a higher $\alpha$ is required to match the
statistics of the observation.
If we  require agreement with method 1, then $\beta = 0$ and $\alpha = 2.2$, compatible
with the absence of a detectable trend in the decay times. Even without constraining method
2, the lowest possible $\alpha$ range (for the lowest acceptable $\beta = 0$)
is around 2.2, i.e., values below $2$ remain excluded.

\section{DISCUSSION}

\subsection{The Validity of Power Laws}

The present analysis was primarily motivated by the  finding that the occurrence frequencies of 
solar flare energies are distributed in power laws. Although some statistical models support
such distributions (e.g., \citealt{lu91,vlahos95}) their physical cause is a matter of debate.
Indeed, there is little reason to assume that the same power laws  strictly hold for all classes
of flares and  for all coronal regions (active regions, magnetic areas above the network,
quiet regions, etc). A simple estimate, for example, suggests the presence of a high-energy cut-off:
An average coronal magnetic field of strength 100 G in a half-spherical active region
of radius $10^{10}$ cm contains $8\times 10^{32}$ erg of magnetic energy. Stronger flares
(as often observed in active stars) either require stronger magnetic fields or a larger volume,
but both are constrained by the maximum magnetic field strength available in the photosphere
and, respectively,  by the volume of active regions (that are related to the size of bipolar magnetic
regions in the photosphere and by about two coronal scale heights; \citealt{serio81}).
A high-energy cut-off is also suggested from avalanche model simulations \citep{lu91,lu93}
and evidence for its existence has been found in observations of small solar active regions
\citep{kucera97}, and in investigations of time-dependent  $\alpha$ values for solar flares 
\citep{bai93,bromund95}.
On the low-energy side,  the power-law index may change due to physical changes in
the energy release \citep{hudson91} and this seems to be the case for solar microflares
as opposed to the  larger flares \citep{krucker98}. A low-energy cut-off is also required 
to confine the total radiated power  if $\alpha \ge 2$ (see \S 1).

We do not construct the flare distributions explicitly with our methods, but
use model light curves for statistical comparison. It is, however, clear that single power laws
describe the observed light curves acceptably well. Although flare energies
detected  by these observations  are comparable to medium-to-large solar
coronal flares, we find $\alpha$ values that  are markedly larger than those reported
for equivalent solar flares, but that resemble those recently reported for solar
microflares with  $\approx 10^6$ times smaller energies. Typical activity indicators
for active stars such as ours (like $L_{\rm X}/L_{\rm bol}$, or the surface X-ray flux) are 
of order 1000 times larger than the Sun's. It thus appears that the regime of steep power-law indices
($\alpha > 2$)  is shifted upwards in energy on active stars, and that we see an
equivalent population of flares at larger energies. We find power-law indices around 2.0--2.5 and
our methods exclude values below $\alpha = 2$ for the softer DS and LECS light curves.
Furthermore, the smallest flare energies that we require to fully model the observations lie 
considerably below the flare detection limit in the binned light curves (Table~\ref{minenergy}) although our second
method may reach a sensitivity close to such values \citep{kashyap02}.   
Like the solar microflares, our population of flares may thus be sufficient to energize the complete  
corona, including the quiescent emission, if the power-law is extrapolated to flares with 
radiated energies 
of a few times $10^{29}$~erg  (Table~\ref{minenergy}). Such energies correspond to relatively 
moderate flares in the solar context.   These conclusions fully support the findings by 
\citet{audard99, audard00}.

\subsection{The Minimum Flare Energies and Quiescent Emission}\label{quiescent}
  
A lower cut-off to the flare energies is required if $\alpha \ge 2$. The cut-off
does not imply that lower-energetic flares do not exist. But it implies 
that  their occurrence rate cannot
follow the extrapolation of the power-law found at higher energies but must be
considerably smaller, effectively introducing a cut-off below which flares 
contribute little. Our simulated model light curves were  
calculated assuming such an energy cut-off. We determined  the cut-off
energy after renormalization and thus found the minimum number of counts
of any of the simulated flares. The values are reported in Table~\ref{minenergy}, column 4.
To convert the total number of counts to total (radiated) flare energy, we computed
the count-to-energy conversion factor as follows: A 3-temperature model (reported in Table~\ref{saxfit}) 
determined from the LECS and MECS data by spectral fitting of the quiescent emission 
was convolved with the response matrices of  each of the three instruments. 
We thus obtained the expected total count rates in the detectors, and by integrating 
the complete spectrum from 0.01~keV to 50~keV we computed the total coronal luminosity. 
We thus find that one detected (quiescent) count corresponds to 
$2.8\times 10^{29}$~erg for the {\it EUVE} DS, to $4.8\times 10^{29}$~erg for the {\it BeppoSAX}
LECS, and to $2.6\times 10^{30}$~erg for the MECS. From this, we obtain the minimum
energy of flares used for the simulation, as shown in column 5 of Table~\ref{minenergy}.  
In other words, extrapolating the power-law distribution of flares down to the reported energies 
is necessary and sufficient to explain the complete observed quiescent flux. The 
LECS provides a lower limit of $4\times 10^{29}$~erg.
For the derived flare time profile, this energy corresponds to a peak luminosity of
approximately $1\times 10^{27}$~erg~s$^{-1}$ in the combined X-ray and EUV ranges.
Such flares  correspond to small solar flares.
Similar values hold for the MECS and DS data (Table~\ref{minenergy}, columns 5 and 6).

We note that the actually {\it detected} flares in the binned light curves
(method 1) typically exceed these levels
by $\sim$2 orders of magnitude. A small flare reaching a peak count rate 3$\sigma$ above 
its pre-flare level corresponds to $3\times 10^{31}$~erg in the DS, $1.4\times 10^{31}$~erg
in the LECS, and $3.6\times 10^{31}$~erg in the MECS (Table~\ref{minenergy}, column 7). The apparently 
quiescent level is thus composed of flares between the cut-off limit and the 3$\sigma$ 
count rate detection limits. In terms of peak luminosity, this interval
covers the $\sim 5\times 10^{25} - 10^{29}$~erg~s$^{-1}$ range (see Table~\ref{minenergy}, columns 6 and 8, 
for details).
Assuming an average radiative loss function (radiative energy loss per unit EM) of 
$\Lambda = 2\times 10^{-23}$~erg~cm$^3$~s$^{-1}$ (appropriate for $T = 5-40$~MK)
we find that the peak EMs of these flares are approximately $2.4\times 10^{48}-5\times 10^{50}$~cm$^{-3}$
for the DS quiescent level, $5\times 10^{49}-2\times 10^{51}$~cm$^{-3}$ for the LECS quiescent
level, and $1.5\times 10^{50}-5\times 10^{51}$~cm$^{-3}$ for the MECS quiescent level.
From Figure 2 of \citet{feldman95} we estimate that the peak  temperatures of these flares
are 15--30~MK for the DS, 25--33~MK for the LECS, and 27--37~MK for the MECS (after
\citealt{aschwanden99}, these temperatures are smaller by a factor of $\sim 1.5$).  If we consider
that most of the flares in the power-law distribution are close to the lower end of the
temperature intervals, we see that the LECS quiescent level is primarily composed of flares that reach 
no more than about 20~MK at peak but mostly reside at lower temperatures.  The MECS effective
area shows a steep gradient below 25$-$30~MK, thus most of the LECS quiescent emission is
suppressed in the MECS. The 3-$T$ fit explicitly shows that the bulk quiescent plasma is at 
temperatures around 10~MK (Table~\ref{saxfit}).

\begin{figure} 
\epsscale{1.0}
\plottwo{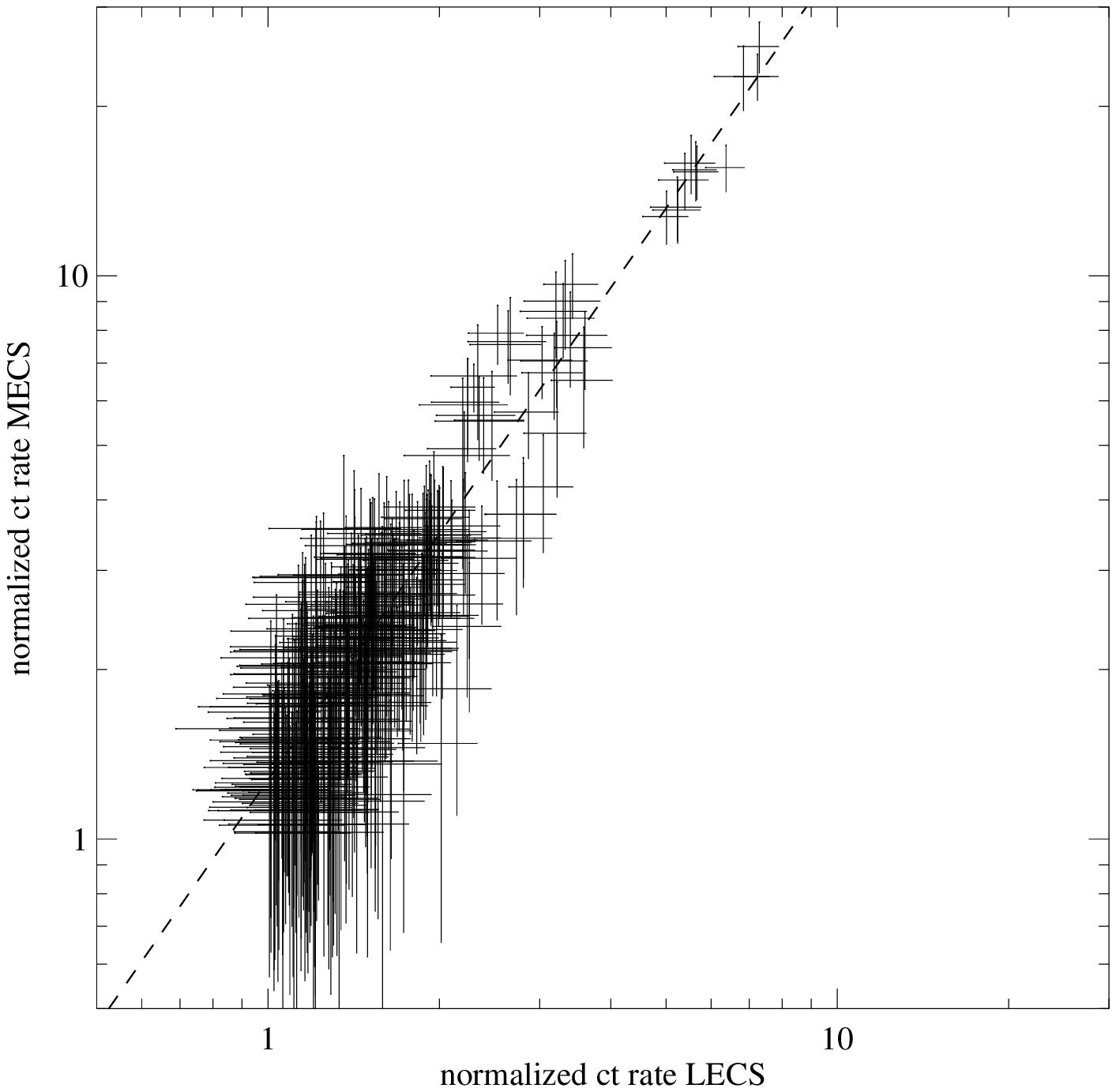}{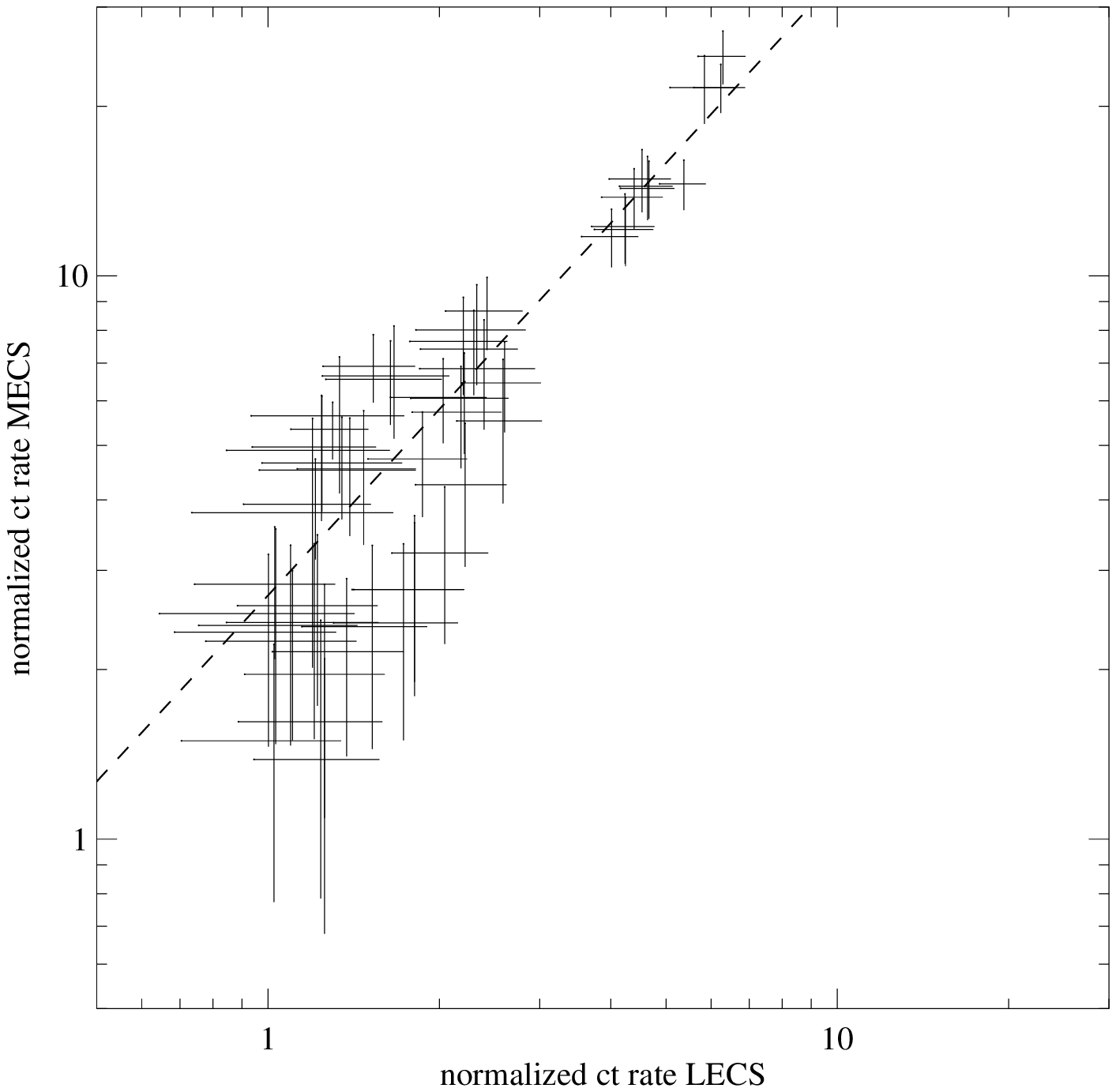}
\caption{Correlation between simultaneous LECS and MECS count rates measured in
 658 bins of 200~s duration each, for the complete light curve. The count rates were
 normalized with the quiescent values, and only  normalized values exceeding 
 unity are plotted. {\bf Left:} All data included. Best-fit slope 
of the regression fit (dashed, using weights calculated from error bars): $1.45\pm 0.02$.
-- {\bf Right:} Quiescent count rate level was subtracted before normalization. Best-fit slope 
of the regression fit (dashed): $1.10^{+0.06}_{-0.05}$.\label{figure8}}
\end{figure}

The two {\it BeppoSAX} light curves indeed look qualitatively different 
(Fig.~\ref{figure2}). Although the same flares are present in both light curves, the 
LECS shows appreciable quiescent emission while the MECS does not. The contrast 
between strong flares and low-level emission is much stronger in the latter. 
This difference is a  consequence of different temperature
sensitivities of the detectors, the MECS being insensitive to low-level
emission from presumably cooler plasma. 
To quantify this hardening effect, the count rates were first normalized 
to the average quiescent count rate level. Fig.~\ref{figure8} (left) illustrates the
relation between LECS and simultaneous MECS count rates for 658 bins of 200~s each (see 
Fig.~\ref{figure2}) along the complete {\it BeppoSAX} light curve.
It shows a pronounced deviation 
from proportionality: the harder MECS emission increases faster than the corresponding softer
LECS emission, i.e., the emission hardens with increasing overall count rate. The best-fit power-law
in this figure  has an index  of $1.45\pm 0.02$.

We thus find a natural explanation for the largely differing contrast between flares and quiescent emission
in the BeppoSAX detectors (Figure~\ref{figure8} left) {\it in the framework of the stochastic-flare heating
hypothesis}. We next need to investigate whether the larger flares themselves 
show  a trend for hardening with increasing count rate. We emphasize again that we measure
hardening by using two different effective area curves for two detectors, not by applying
any spectral analysis.

There are two possible contributions that induce the non-proportionality between normalized LECS and
MECS count rates in Fig.~\ref{figure8} (left):
i) There is an intrinsic hardening  for the larger flares, as suggested by the Feldman et al.
    and {Asch}\-wan\-den et al.
   relations. Thus, if the detector effective area curves vary differently
   across the relevant temperature range for LECS and MECS, we expect to see hardening signatures,
   i.e., non-proportionalities in Fig.~\ref{figure8}.   
ii) The two detectors have, as argued above, different relative sensitivities
   to the quiescent emission compared to the (presumably hotter) large flares.  To 
   compare the flare count rates, the quiescent levels should be subtracted.
   We thus subtract a baseline quiescent count rate ($0.105 \pm 0.010$~cts~s$^{-1}$
   for the LECS and $0.025 \pm 0.005$~cts~s$^{-1}$  for the MECS). Fig.~\ref{figure8} (right)
   shows the relation between the residual count rates for LECS and MECS after the optimum subtraction.
   The slope of the best-fit line is  $1.10^{+0.06}_{-0.05}$, i.e., there is still some hardening
   effect in the flare count rates.
 We show in the next subsection how the hardening of larger flares can
 explain the shallower flare energy distribution found for the MECS (compared to the LECS results).


\subsection{Dependence on Spectral Range}\label{dependence}

The range of best-fit $\alpha$ agrees for all tests performed on
the data from the {\it EUVE} DS and from the {\it BeppoSAX} LECS ($2.1 \la  \alpha \la  2.4$).
There seems to be a systematic shift of the MECS results relative to those 
from LECS, by approximately $\Delta\alpha = 0.2-0.4$ to lower $\alpha$, i.e., the MECS 
distribution is shallower although this effect is only marginally
significant. 

For an isothermal plasma, a harder energy spectrum implies a higher plasma temperature. 
\citet{feldman95} reported a correlation for solar (plus a few stellar) flares
between flare peak temperature $T_0$ and  flare peak emission measure $EM_0$ (which scales approximately 
with the flare peak luminosity and, for constant decay times, with the total radiated 
flare energy). The relation is nearly exponential 
but can be reasonably approximated also by a power law over a limited range of 
temperatures,
\begin{equation}\label{e:feldman}
\mathrm{EM}_0  = aT_0^b \quad {\rm [cm^{-3}]}
\end{equation}
with $a\approx 2\times 10^{13}$~cm$^{-3}$K$^{-b}$, $b\approx 5\pm 1$ in the range of $T = 5-30$~MK, 
and $T$ measured in K. \cite{aschwanden99} reports a power-law dependence
between $T = 1-20$~MK with $b\approx 7$. 
The X-ray luminosity in general can be expressed as 
\begin{equation}\label{e:luminosity}
L \approx \mathrm{EM}~\Lambda(T) = f~\mathrm{EM}~T^{-\phi}    
\end{equation}
with $\phi \approx 0.3$ over the  above temperature range (for broad-band X-ray losses
as derived in XSPEC or SPEX).

The count rate measured in a detector depends on the effective area curve and the 
incident X-ray spectrum which in turn depends on the plasma temperature $T$ and the 
emission measure. We approximate the dominant X-ray emission at any one time as 
being emitted by an isothermal plasma of temperature $T$, with equation~\ref{e:feldman} 
satisfied, i.e., we identify any count rate at any given time with an individual  
flare at its peak and thus use its peak $EM = EM_0$ and peak $T = T_0$ for this estimate.
Because we cannot perform time-resolved temperature analysis
with the present data, we approximate the true flare energy distribution with
the observed count distributions (i.e., $c\propto E$; the indices ${\ell}$ and $m$ stand
for LECS and MECS, respectively):
\begin{eqnarray}\label{e:powerlaws}
{dN\over dc_{\ell}} &\propto& c_{\ell}^{-\alpha_{\ell}} \\
{dN\over dc_{m}} &\propto& c_{m}^{-\alpha_{m}}
\end{eqnarray}
where the right-hand sides  hold for the hypothesis that the energy
distributions (i.e., the count rate distributions) are power laws. This approximation is
acceptably good for outstanding flares, i.e., count rates well above the quiescent level,
 but breaks down near the quiescent level where numerous small flares may overlap. 

We describe the dependence of the LECS and MECS count rates $c_{\ell}$ 
and $c_m$  on temperature $T$ and luminosity $L$ with power-law approximations 
\begin{eqnarray}
c_{\ell} &=& c_{\ell,0} L T^{\gamma_{\ell}} \\
c_{m} &=& c_{m,0} L T^{\gamma_{m}} \label{powercount}
\end{eqnarray}
where $c_{\ell,0}$ and $c_{m,0}$  are detector-related constants and
$\gamma_{\ell}$ and $\gamma_{m}$ describe the temperature sensitivity of the detector 
count rate. The $\gamma$ values must be determined from the detector effective areas. 
The expression ${\cal{E}}_{\ell} = c_{\ell}/L = c_{\ell,0}T^{\gamma_{\ell}}$ 
(similar for $m$) gives the observed {\it count rate per unit luminosity}, i.e., 
the efficiency of the detector to record a given luminosity in terms of a count 
rate, as a function of temperature. 

\begin{figure}
\plotone{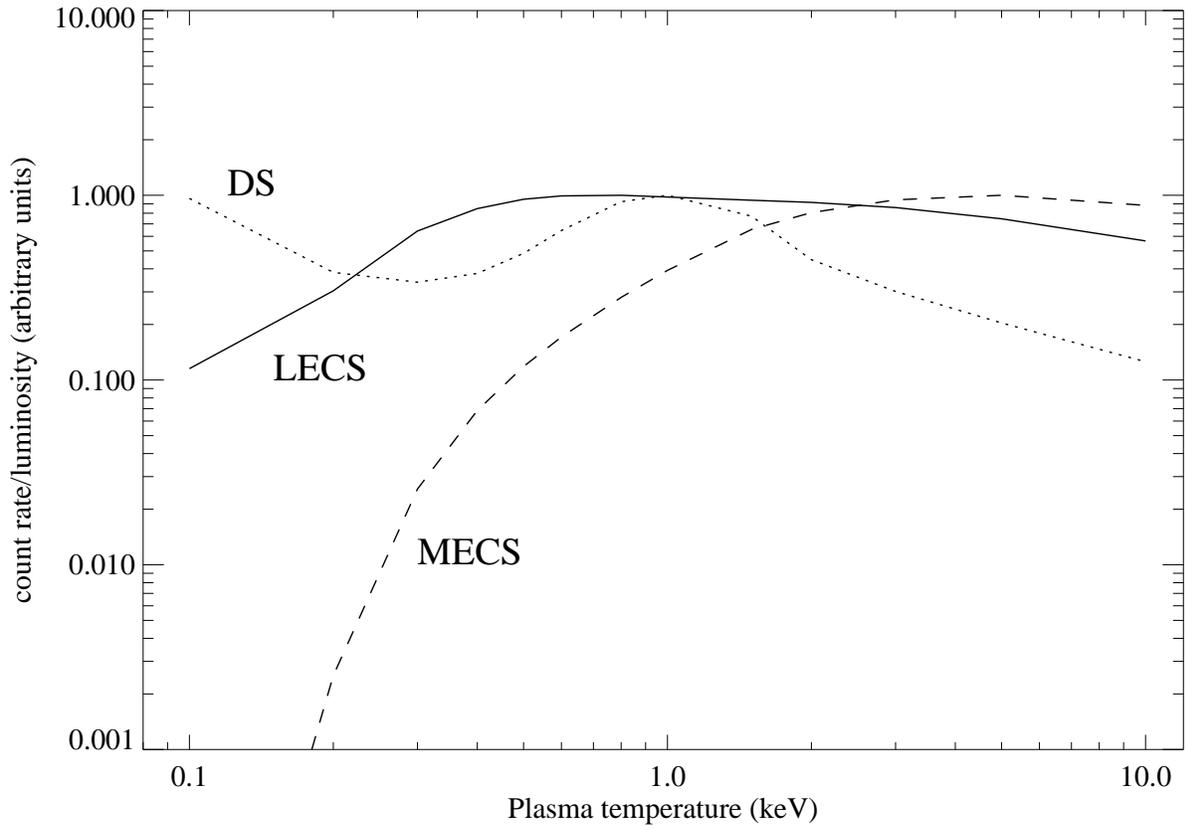}
\caption{Efficiency (count rate/luminosity) as a function of (isothermal) plasma temperature 
      for the {\it EUVE} DS, and the {\it BeppoSAX} LECS and MECS detectors (arbitrarily normalized).
      \label{figure9}}
\end{figure}

We have folded a number of theoretical isothermal model spectra (in
collisional equilibrium, as modeled in XSPEC [\citealt{arnaud96}],
and SPEX [\citealt{kaastra96a}]) with the detector response matrices 
for the DS, LECS, and MECS detectors to describe ${\cal{E}}_{\ell,m}(T)$. The count rates
were derived across the range of sensitivity of the detector, and the
$L$ values have been evaluated for each $T$ by integration of the model spectrum from 
0.01$-$50~keV  assuming  a unit emission measure. 
The results are shown in Fig.~\ref{figure9}. The efficiencies of the LECS and DS detectors are 
rather flat ($\gamma_{\ell} \approx 0$) at least in the region of interest, i.e., 
within approximately $0.5-5$~keV. The LECS is least sensitive to changes in $T$ and 
therefore {\it best recognizes flares of different temperatures with the least bias.} 
On the other hand, the efficiency of the MECS is a strong function of $T$ below 
$\sim 3$~keV, with a rapid drop-off toward emission from cooler plasmas. Because 
of the dependence in equation~\ref{powercount}, this implies that {\it weak flares are suppressed 
in the MECS, and strong flares are enhanced}. This is equivalent to an apparent decrease
of $\alpha$ compared to the  initial energy distribution. 

To estimate this effect, we set $\gamma_{\ell} \equiv 0$ (for the temperature range of 
interest). 
The number distribution of flares (i.e., counts) in LECS count rate, 
$dN/dc_{\ell}$, is therefore an unbiased approximation to 
the true flare radiative energy distribution $dN/dE$ since $E \propto L$.  Since
\begin{equation}
{ c_{m}\over c_{\ell}} = {c_{m,0}\over c_{\ell,0}}T^{\gamma_m} 
                            = {c_{m,0}\over c_{\ell,0}}
			      \left({L\over af}\right)^{\gamma_{m}/(b-\phi)} 
                            = {c_{m,0}\over c_{\ell,0}}
			      \left({c_{\ell}\over afc_{\ell,0}}\right)^{\gamma_{m}/(b-\phi)}
\end{equation}
(using equations~\ref{e:feldman} and \ref{e:luminosity})  we obtain 
\begin{equation}
c_{m} \propto c_{\ell}^{\gamma_{m}/(b-\phi) + 1}.
\end{equation}
Therefore,
\begin{equation}
 {dN\over dc_{\ell}} \equiv {dN\over dc_{m}}
 {dc_{m}   \over dc_{\ell}} \propto
 {dN\over dc_{m}} c_{\ell}^{\gamma_{m}/(b-\phi)}.
\end{equation}
With equations~\ref{e:powerlaws} and 6 we find that
\begin{equation}
  \alpha_{\ell} = \alpha_{m}\left( {\gamma_{m}\over b-\phi} + 1 \right) 
         - {\gamma_{m}\over b-\phi}.
\end{equation}
According to \S~\ref{quiescent}, two components may be responsible for the apparent decrease
of $\alpha$; we consider them to be limiting cases. 
(i) The hardening in Fig.~\ref{figure8} (left) could be attributed to the intrinsic 
spectral hardening toward larger energies of the visible flares only. 
We found $c_{m} \approx c_{\ell}^{1.45\pm 0.02}$, hence  
$\gamma_{m}/(b-\phi) + 1 = 1.45$ and therefore $\gamma_{m}/(b-\phi) = 0.45$, so that
$\alpha_{\ell} = 1.45\alpha_{m} - 0.45$. We find the highest confidence for
$\alpha_{\ell} \approx 2.4$, hence we expect high confidence for 
$\alpha_{m} \approx  1.97\pm 0.02$ (not including possible systematic errors of $\pm 0.1$ as
discussed above).

ii) Preceding subtraction of the quiescent count rate level both for LECS and MECS suggests 
(Figure~\ref{figure8} right)  $c_{m} \approx c_{\ell}^{1.10(+0.08,-0.06)}$, hence  
$\gamma_{m}/(b-\phi) + 1 = 1.10$ and $\gamma_{m}/(b-\phi) = 0.1$, so that
$\alpha_{\ell} = 1.10\alpha_{m} - 0.10$. We thus  expect high confidence for 
$\alpha_{m} \approx  2.27\pm 0.08$.

The expected difference $\alpha_{l}-\alpha_{m} \approx 0.13-0.43$ agrees with the
measured difference of $0.2-0.4$. We conclude that the hardening of flares
with increasing total energy significantly affects the determination of $\alpha$.
While the LECS data provide, thanks to the
flat efficiency curve, a relatively unbiased measure of the flare energy distribution
(in terms of emitted energy), the MECS data consistently result in a harder appearance of
the light curve, especially  if the quiescent emission is included.
We thus recognize the MECS result as biased by detector properties. 
As far as the DS results are concerned, they agree well with the LECS results, which is 
consistent with the relatively flat efficiency curve of the DS (Fig.~\ref{figure9}).

Similar temperature bias has been discussed for solar coronal flare statistics derived
from emission line flux ratios in the EUV range covering a narrow temperature range 
\citep{aschwanden02a, aschwanden02b}.
One may extend this analysis to solar HXR data that have traditionally been used
as a diagnostic of the total flare energy release \citep{lin84}. If the production
of hard X-rays is more efficient in strong flares (the ``Big Flare Syndrome'', Kahler 1982),
then hard X-ray flare energy distributions are shallower than those constructed from
soft X-ray or EUV data. There are at least indications that very small, low-temperature
flares on the Sun are poor in radio emission, i.e., they produce high-energy
electrons (important for the HXR production) less efficiently than larger flares \citep{krucker00}.

\subsection{The Emission Measure Distribution}\label{dem}

We will now investigate effects of our hypothesis in the limiting case that
all of the observed X-ray and EUV emission is due to superimposed stochastic-flare emission. 
In that case, we can use our $\alpha$ values to develop a crude
model for the time-averaged EM distribution of the observed corona.
We will estimate the amount of  EM at a given temperature produced
by  a flare of any energy during its decay phase, and then integrate over the 
flare energy distribution. The time a flare spends at a given temperature is used
as a weighting factor.

\begin{figure}
\plotone{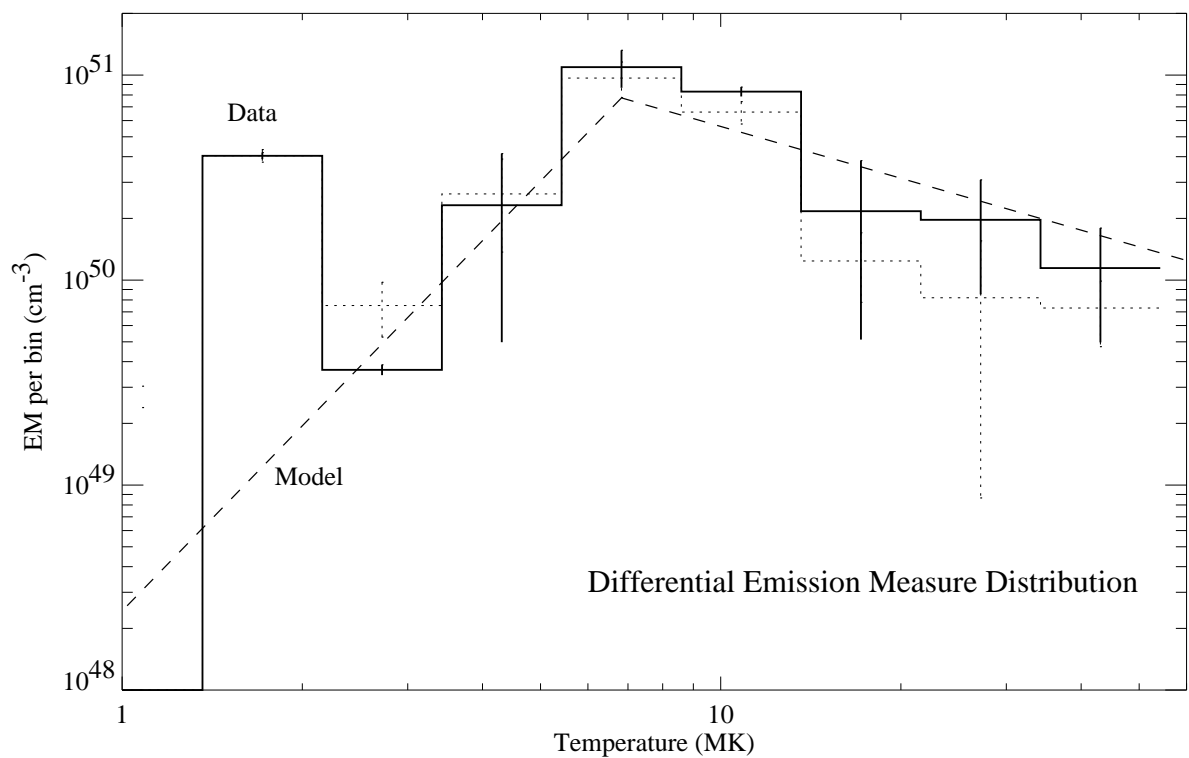}
\caption{Emission measure distribution of the quiescent emission of AD Leo, derived
    from the combined {\it BeppoSAX} LECS and MECS data. The bins are equidistant in
    $d\mathrm{log}T$. The emission measures
    are averages derived from a polynomial fit  and a regularization method
    (for details, see \citealt{kaastra96b}). The dashed lines illustrate  the approximate
     slopes of the DEM expected from a superposition of flares (arbitrarily 
     adjusted in EM$_{\mathrm{peak}}$).\label{figure10}}
\end{figure}

From equations~\ref{e:feldman} and \ref{e:luminosity} we obtain a relation between
the flare peak temperature $T_0$ and its peak luminosity $L_0$,
\begin{equation}\label{e:peakt}
T_0 = \left({L_0\over af}\right)^{1/(b-\phi)}.
\end{equation} 
Note that for  decay times $\tau$ independent of $E$, the peak luminosity follows
the same power-law as the total radiated energy:
\begin{equation}\label{e:peakdistrib}
{dN\over dL_0} = k\tau^{-\alpha+1} L_0^{-\alpha} 
\end{equation}
(since $E = L_0\tau$). However, we will also investigate the case in which $\tau$ varies with the
flare energy, namely
\begin{equation}
\tau = \tau_{\rm 0} E^{\beta}
\end{equation}
(where $\beta \ge 0$ is assumed, and $\tau_{\rm 0}$ is a constant adjusted to the larger 
detected  flares) in which case
\begin{equation}\label{e:peakdistribtau}
{dN\over dL_0} ={dN\over dE}{dE\over dL_0} = k^{\prime}L_0^{-(\alpha-\beta)/(1-\beta)}.
\end{equation}
where the constant $k^{\prime} = k\tau_{\rm 0}^{(1-\alpha)/(1-\beta)}/(1-\beta) > 0$ as 
long as $\beta <1$ (which can be reasonably assumed).
Since we neglect the short rise time of the flare, our flare light curves are
described  by their  exponential decay at $t \ge 0$,
\begin{equation}\label{e:decay}
L(t) = L_0e^{-t/\tau}.
\end{equation}
From hydrodynamic modeling, theory and observations, it is known  
that during the flare decays $T\propto n^{\zeta}$, where $n$ is the plasma density
\citep{reale93}.
The parameter $\zeta$ is usually found between 0.5 and 2  \citep{reale93}.
A value of $\zeta = 2$ holds if the heating source   abruptly turns
off at flare peak, and the flare cools freely via radiation and conduction.
A low value of $\zeta$ indicates sustained heating during the decay phase.
For example, for $\zeta \approx 0.5 - 0.7$ the decay time of the 
flare emission is 2$-$4 times slower than predicted for a freely cooling
magnetic loop \citep{reale97}. Solar observations  of moderate flares  with {\it SMM}
show  $\zeta$ between $\sim$0.5--2 \citep{sylwester93}, although \citet{reale97} find
a predominance of values around $\sim 0.3 - 0.7$, i.e., flares with sustained heating.
For larger stellar flares, there is much evidence for small $\zeta$ as well:
\citet{reale98} find a flare decay that is $\ga 2$ times slower than for a freely cooling 
loop; \citet{favata00a} find $\zeta = 0.56\pm 0.04$ for a flare on EV Lac, and
\citet{favata00b} report $\zeta = 0.48\pm 0.06$ for a flare on AD Leo; \citet{gudel01b}
find $\zeta = 0.95\pm 0.15$ for a moderate flare on AB Dor.  
Since, under the assumption of constant volume, the emission measure scales
with $n^2$, we have 
\begin{equation}\label{e:scaling}
{T\over T_0} = \left({\mathrm{EM}_T\over \mathrm{EM}_0}\right)^{\zeta/2}.
\end{equation}
From equations~\ref{e:luminosity}, \ref{e:decay} and \ref{e:scaling}, we obtain the temperature
time evolution
\begin{equation}\label{e:tempdecay}
T(t) = T_0e^{-t/(2\tau[1/\zeta - \phi/2])} 
\end{equation}
where we require $\phi < 2/\zeta$ for a temperature decay (this is usually fulfilled
as $\phi \approx 0$ and $2/\zeta \ge 1$).
The emission measure $\mathrm{EM}_T$ at temperature $T$ for this particular
flare is, from equations~\ref{e:scaling} and \ref{e:feldman}, 
\begin{equation}\label{e:emt}
\mathrm{EM}_T = aT_0^{b-2/\zeta}T^{2/\zeta}
\end{equation}
and this expression must be weighted with the fractional duration (out of 
the total observing time $P$) during which
the flare resides within the logarithmic temperature interval 
$[\mathrm{ln}T, \mathrm{ln}T + d\mathrm{ln}T]$, i.e., $dt/P$. The emission measure
contribution $\mathrm{EM}_T$ 
thus counts in full if it is constantly present during the observing time.
The duration $dt$ is obtained 
by differentiating equation~\ref{e:tempdecay} (we only require the absolute value of
$d\mathrm{ln}T$), so that the weight is
\begin{equation}\label{e:deltat}
{dt\over P} = {2\tau\over P}\left({1\over \zeta}-{\phi\over 2}\right)|d\mathrm{ln}T|.
\end{equation}
Multiplying equations~\ref{e:emt} and \ref{e:deltat} and substituting all expressions, 
we find the weighted contribution of a  flare
with peak luminosity $L_0$ to the emission measure at temperature $T$, 
\begin{eqnarray}\label{e:deltaem}
\delta(\mathrm{EM}_T) &=& {2a\tau_{\rm 0}^{1/(1-\beta)}\over P} \left({1\over \zeta}-{\phi\over 2}\right) \nonumber\\ 
         &\times& \left({L_0\over af}\right)^{(b-2/\zeta)/(b-\phi)}L_0^{\beta/(1-\beta)}T^{2/\zeta}|d\mathrm{ln}T| \nonumber\\
               &=& cL_0^{(b-2/\zeta)/(b-\phi) + \beta/(1-\beta)}T^{2/\zeta}|d\mathrm{ln}T|.
\end{eqnarray}
where all constant factors are absorbed in $c$. It is positive as long
as $\phi < 2/\zeta$ (from equation~\ref{e:deltat} and as required for equation~\ref{e:tempdecay}).
Expression~\ref{e:deltaem} must be integrated over $N$ in the distribution~\ref{e:peakdistribtau} 
for all possible peak luminosities $L_0$ to
obtain the differential emission measure distribution (DEM), i.e.,
\begin{eqnarray}\label{e:DEM}
Q(T) &=& {d(\mathrm{EM})\over |d\mathrm{ln}T|} = \int_{L_1}^{L_2}{\delta(\mathrm{EM}_T)\over |d\mathrm{ln}T|}{dN\over dL_0}dL_0 \nonumber \\
  &=& ck^{\prime}T^{2/\zeta}\int_{L_1}^{L_2}
           L_0^{(b-2/\zeta)/(b-\phi) - (\alpha - 2\beta)/(1-\beta)}dL_0\nonumber \\
  &=& \left.{c^{\prime}T^{2/\zeta}\over \mathrm{(exponent)}}
	   L_0^{(b-2/\zeta)/(b-\phi) -(\alpha-2\beta)/(1-\beta) + 1}\right|_{L_1}^{L_2} 
\end{eqnarray}
where $c^{\prime} = ck^{\prime}$ is again a numerical constant that is $> 0$ if $\beta < 1$ (which is fulfilled; also
note that $\beta \neq 1$ is required; see equation~\ref{e:peakdistribtau}). The denominator ``(exponent)''
is identical to the exponent of $L_0$ which we assume is negative so that large but rare flares do not
dominate the average emission measure distribution. This holds for reasonable 
choices of $b$, $\beta$, and $\phi$ for a given $\alpha$ but needs to be checked in
individual cases. Then, the upper integration limit can be set to $L_2 = \infty$ as the 
corresponding term converges to zero, and the complete expression~\ref{e:DEM}
remains positive. For the lower limit $L_1$, there are two cases. i) For a given
$T$, there is a smallest flare that  reaches this temperature at its
peak, i.e., $T = T_0$, and this flare then has a peak luminosity given by 
equation~\ref{e:peakt}, i.e., $L_0 = afT_0^{b-\phi}$. ii) Since $\alpha > 2$, a lower energy
cut-off of the flare energy distribution is required. Flares with energies below this 
limit $E_{\mathrm{min}}$ are assumed to be unimportant for the heating.  The minimum radiated 
flare energy corresponds to a minimum peak luminosity $L_{\mathrm{min}} = E_{\mathrm{min}}/\tau$ 
and was reported in Table~\ref{minenergy}. We therefore require
\begin{equation}
L_1(T) = \mathrm{max}(afT^{b-\phi}, L_{\mathrm{min}}).
\end{equation}
Consequently, there are two temperature regimes, depending on whether the first or second expression applies. From
equation~\ref{e:DEM}, we obtain
\begin{eqnarray}
Q &\propto& T^{2/\zeta}\quad\quad\quad\quad\quad\quad\quad\quad\quad\   , afT^{b-\phi} \le  L_{\mathrm{min}} \\
                                    &\propto& T^{-(b-\phi)(\alpha-2\beta)/(1-\beta) +2b - \phi} \quad       , afT^{b-\phi} >  L_{\mathrm{min}}
\end{eqnarray}
The EM distribution thus rises with a slope of $2/\zeta$ from low temperatures toward a turnover, and 
then drops again with a power-law exponent that depends on the flare energy distribution index $\alpha$. 
The turnover occurs where $afT^{b-\phi} = L_{\mathrm{min}}$. Note that for $\zeta = 0.3-0.7$ \citep{reale97},
$Q \propto T^{3-7}$ which is much steeper than the prediction from a quasi-static loop model but
agrees well with measurements of the DEM of the active K star $\epsilon$ Eri \citep{laming96} and
of the late G star $\xi$ Boo A \citep{drake01a}. The DEM thus provides another important
diagnostic for flare-heated coronae:
i) The high-$T$ slope of the distribution can be used to determine $\alpha$; ii) the low-$T$ slope gives
information on the heating time-scale of flares during their decay; and iii) the peak indicates
the turnover energy from a steep to a shallow flare energy power-law. --
We briefly discuss two cases, for which we use $\alpha = 2.2$.

i) For $\tau$ independent of flare energy, $\beta = 0$. 
Our simulations required lowest energies corresponding to peak luminosities as low as a few times  
$10^{25}$~erg~s$^{-1}$; we assume $L_{\mathrm{min}} = 10^{25}-10^{26}$~erg~s$^{-1}$.  
The cooling losses per unit emission measure are only weakly dependent on temperature
in the X-ray range (0.1--10~keV), namely $\Lambda \approx 2\times 10^{-23}$~erg~cm$^3$~s$^{-1}$ 
(determined from a MEKAL model in XSPEC using our best-fit abundance for Fe).
The peak EM required for such a flare  is thus  $5\times 10^{47}- 5\times 10^{48}$~cm$^{-3}$,
which, from the Feldman et al. relation, corresponds to a temperature of about 9$-$16~MK, with 
considerable scatter. This is indeed the temperature regime of peak EM in the emission measure 
distribution. It is, however, 
unlikely that the flare energy distribution is limited by an abrupt cut-off at low energies.
Rather, the power-law index is likely to become smaller at lower energies so that smaller flares 
become less relevant for heating. We therefore set the turnover temperature
arbitrarily at $\sim$7~MK, in agreement with the observational DEM. The high-$T$ part of the DEM 
is thus determined mainly by plasma between 10$-$40~MK (see Figure~\ref{figure10}). 
The Feldman et al. relation is
best fitted by $b = 6\pm 1$ in that regime, and the cooling function (as calculated in XSPEC for
an Fe abundance of 0.54 times solar photospheric) has an average slope of $\phi = 0.3$ in that
range. We propose to set $\zeta = 0.5 - 1$, 
in the light of the above discussion of published values. We thus find
$Q  \propto T^{3\pm 1}$ for the low-$T$ part of the emission
measure distribution, and 
$Q \propto T^{b(2-\alpha)-\phi(1-\alpha)} = T^{-0.2b+0.36} = T^{-0.84\pm 0.20}$ for the high-$T$ part.
This model DEM is shown in Fig.~\ref{figure10} (dashed) together
with a quiescent DEM (dotted) and a DEM from all data (solid), both
derived from LECS and MECS (using the polynomial and the 
regularization method as described in \citealt{kaastra96a}). Clearly, including the visible flares
fills in more emission measure at temperatures between 20--50~MK, as expected. This explicitly
confirms that the high-$T$ DEM is largely related to flaring emission. The fact that the quiescent
DEM in Figure~\ref{figure10} still contains plasma at $\sim$40~MK is likely to be due to the
inclusion of many small flare peaks not cut out for this fit.  
The reconstructed DEM has limited quality given the restricted resolution of the {\it BeppoSAX}
detectors, but the comparison with the theoretical DEMs is suggestive except for the presence
of a cool peak at 0.15~keV.

ii) We previously discussed the possibility that $\beta = 0.25$. Again, we set $\phi = 0.3$ and
$\zeta = 0.5 - 1$. Then, the low-$T$ slope remains the same, $3\pm 1$, and the high-$T$
DEM varies like  $Q \propto T^{-0.267b + 0.38} =  T^{-1.22\pm 0.27}$,
i.e., quite similar to case (i).

\section{Conclusions}

We have investigated the role of statistical flares in coronal heating of magnetically active
stars. Long observations of AD Leo were obtained in order to maximize flare statistics. 
Flares have been suspected to
play an important role in coronal energy release and subsequent impulsive heating of
chromospheric material to high temperatures.  Chromospheric evaporation
induced by chromospheric overpressure lifts the hot plasma into the corona where it
fills closed magnetic loops. Since (solar) flares are always related not only to
an increase in emission measure but to a significant increase in the average
plasma temperature, they are natural candidates  to heat perhaps all of the
detected coronal plasma. Recent progress in solar physics \citep{krucker98,
aschwanden00, parnell00} has added new momentum to this hypothesis.

Active (but quiescent) stellar coronae exhibit a number of features unknown to the non-flaring Sun
but suspiciously reminiscent of solar (or stellar) flares: i) Very high temperatures up to 2$-$3~keV,
similar to temperatures of large solar flares; ii) accompanying, strong non-thermal
gyrosynchrotron radio emission attributed to relativistic electrons accelerated in the initial phase
of the flare energy release (\citealt{gudel94} and references therein) ; iii) high densities 
($\ga 10^{10}$ cm$^{-3}$)
reminiscent of (solar) flare densities \citep{gudel01a,gudel01b}; iv) "anomalous"
elemental abundances  tentatively ascribed to the action of flares, perhaps analogs to solar Ne- 
and S-rich flares \citep{brinkman01, drake01}; v) and finally, the presence of a large number of strong flares, 
where the rate of detected flares
correlates with the quiescent emission  level \citep{audard00}.

We have studied the distribution of EUV and X-ray flares in energy, seeking power laws of the
form $dN/dE =kE^{-\alpha}$ where $k$  is the normalization of the distribution and  $\alpha$ determines
the steepness of the distribution.  We have applied two methods, one based on the count rate distribution 
of binned data, and the second related to the distribution of arrival-time differences of 
the original   photon lists (only for {\it EUVE}). Despite  the fundamentally
different approaches,  the results of both methods are in excellent agreement for the {\it EUVE}
data  and indicate $\alpha = (2.1-2.3) \pm 0.1$. Simultaneous X-ray observations
obtained with the {\it BeppoSAX} LECS were treated with the first method. The results again overlap with the
{\it EUVE} results, namely $\alpha = 2.4\pm 0.2$. Only the MECS data show somewhat shallower
distributions, with $\alpha = (2.0-2.2)\pm 0.2$ which can be explained by the harder sensitivity
range of the MECS detector, and detection bias in terms of flare temperatures. Our results
are compatible with the findings of \citet{audard00} who applied a flare identification algorithm to 
explicitly record flares and to measure their energies. They are further supported by the findings
of \citet{kashyap02} who study further active stars with the {\it EUVE} DS. At first sight, our 
$\alpha$ values  support a model 
in which the complete coronae are heated by a statistical distribution of flares, involving flares with 
energies down to a few times $10^{29}$ erg (of radiated energy). Also, a model EM distribution based
on the superposition of flares of different peak temperatures is compatible with the
observed EM distribution. However, before we  can conclude that flares 
play an important role in the coronal heating process, we should keep in mind the following 
caveats that stellar observations of the present quality are invariably subject to:

(A) All EUV and X-ray  observations refer to the radiated energy  from
the hot plasma. There is considerable  ignorance of other energy partitions that also contribute
to the  flare energy budget  (\citealt{wu86}):  Kinetic energy of the
upstreaming plasma; potential energy of the lifted plasma; energy in waves; energy in accelerated
particles;  and energy released at longer wavelengths. To interpret our results physically, 
we  adopt the following working hypothesis: 
i) The energy initially released in energetic
particles is largely thermalized in the chromospheric evaporation process. ii) The remaining 
energy (kinetic and potential) is eventually thermalized  (e.g., when material
drops back to the chromosphere). iii) All thermal energy is eventually radiated away during
the cooling processes. We emphasize that this applies also to all energy that is conducted 
from the coronal loops downwards. This energy is radiated by the chromosphere.
iv) The fraction of the radiative energy released in the X-ray range is similar for all flares.
 
While points i)--iii) are supported by observations \citep{dennis85}
and by numeric simulations (e.g., \citealt{nagai84, antonucci87}), 
point iv) is relatively difficult to assess. 
Clearly, a considerable part of the energy  is lost at UV wavelengths not accessible to 
our observations. The  tendency of the flare temperature
to increase with overall flare energy \citep{feldman95,aschwanden99} would
suggest that smaller flares lose a larger fraction of their energy outside the EUV/X-ray regime.
This is, however, of little relevance for us:
All flares considered here are quite large, with probable peak temperatures (according to the 
Feldman et al. relation) exceeding 20 MK; even the quiescent emission shows its
peak EM between 5--10 MK (Table~\ref{saxfit}). \citet{hudson91} reports that 
approximately 2/3 of the total radiant energy of a solar flare are emitted in 
soft X-rays.  The total, long-term average energy loss in optical U band 
flares is linearly correlated with the average X-ray losses  in active stars 
\citep{doyle85}. If we missed a
population of very small flares with very low temperatures (e.g., comparable to 
microflares in the Sun), then they would simply add to the flare distribution on the 
low-energy  side, i.e., our distributions would become steeper still, making
small flares even more crucial for the total  energy release.
Further, the measured average thermal energy input
(``heating rate'') during a solar flare scales linearly with the radiative loss rate
in X-rays \citep{aschwanden00}.

Finally, from the phenomenological point of view adopted in the
present study, the relation between losses in the X-rays/EUV  and those at longer 
wavelengths can be ignored  altogether if we keep with our goal of modeling the 
{\it observed coronal} emission. The latter is clearly
dominated by X-ray/EUV radiation. If we successfully  explain the
total X-ray/EUV emission by the radiation  of a statistical ensemble of flares, then this
simply implies that there is no significant additional {\it coronal}
component. Although there may be additional energy release at lower temperatures (i.e., at 
chromospheric  levels), this becomes irrelevant for the question of {\it
coronal} heating.

(B) The power-law distribution of flares may change spatially on the
star. Stellar observations unavoidably 
treat the corona as an average structure. The recent solar results with
$\alpha > 2$ were obtained in regions of the quiet Sun while most larger flares
occur in active regions. It may, however,  be interesting to mention  that an observation 
of a {\it single} solar X-ray bright point resulted in a power-law index similar to 
those obtained from the whole Sun \citep{shimojo99}. 

(C) The power-law distributions may also vary in time. \citet{bai93} and \citet{bromund95} find 
a 154~d periodicity in which $\alpha$ changes by $\sim 0.2-0.4$ in solar data. There may also  
be a dependence on the overall magnetic activity level that varies with the (cyclic or
irregular) ``magnetic activity cycle''. This latter conjecture is, however, not supported
by the solar studies of \citet{feldman97} and \citet{lu91}. We note that we found different
values for $\alpha$ depending on whether or not the early part of the DS observation (containing
a large flare) was included.

(D) Although we have used a rather long observing time series (27.3 days
of coverage with {\it EUVE}, referring to segment I--IV), some chance coincidence, like the very
large flare at the beginning of the observation, may introduce considerable systematic bias. We
have investigated the role of this large flare on the result for $\alpha$ and found indeed that
its selective inclusion/exclusion  can shift the optimum value  by $\Delta\alpha \approx 0.1$.  

(E) There may be high-energy cut-offs (``roll-overs''; \citealt{kucera97}) 
related to the maximum energy that can be liberated
in stellar active regions. In a limited set of observations with a
limited dynamic range (ratio of strongest to weakest detected flares, also depending 
on the noise level), the deficit of flares close to the high-energy cut-off (because
of their small occurrence rate) can induce a steepening of a power law. A consequent
shallower continuation of the distribution toward  flares below our detection limit
would contribute less energy than estimated with our single power-law approach.
The present data do not allow us to judge on the presence or absence of high-energy
cut-offs. The  good representation  by single-power-law flare distributions
does presently not argue for their presence. The continuation of the power law from detected flares
to energies below the detection threshold has been explicitly assumed in our energy
estimates, and this is no different from any previous (solar or stellar) study.
For sufficiently small energies, the large number of small flares involved begin to  
overlap in time (the ``confusion limit'', already evident in
our light curves). They can no longer be measured individually unless
spatial resolution is available.    

(F) Appreciable non-flare contributions to the EUV/X-ray variability are possible  (e.g.,
evolution of non-flaring active regions, newly emerged magnetic regions, 
rotational modulation of active regions). They would  normally add  to the low-level variability 
and may thus tend to steepen the count rate distributions.

Despite these caveats, some of which will be difficult or impossible to avoid in future
observations, we presently see no compelling argument  against our basic finding, 
namely, that flares statistically contribute
an important part to the overall coronal radiative losses, and that they are therefore 
 good candidates for the coronal heating process per se in magnetically active stars.

Our values for $\alpha$ are very similar to those measured for microflares
in the Sun \citep{krucker98, parnell00} despite the 6 orders of magnitude larger
energies involved. This factor in energy may  partly reflect the level of 
magnetic activity. If so, then the role played by microflares in the Sun is played by 
the much larger flares relevant here in active stellar coronae.

We find independent support for flare heating of active stellar coronae in their coronal
emission measure distribution. By statistically co-adding flaring emission measures 
by weighting them with the dwell time at a given temperature, we derived an analytical
expression for the differential emission measure distribution. The DEM is characterized
by a steeply rising  low-temperature part and a falling high-temperature part. 
The slopes
and the turnover temperature are in principle determined by the flare energy power-law 
index $\alpha$, the low-energy break in the distribution, and the flare heating parameter 
$\zeta$ during the flare decay. Previously published DEMs of active stars (e.g., \citealt{laming96, 
drake01a}) and the DEM derived here show characteristic shapes that are compatible with our
expression but are not supported by quasi-static loop models. We suggest that the coronal DEMs 
directly reflect the operation of heating and cooling mechanisms during  
stochastic flares (\citealt{gudel97, gudeleal97}).

We conclude this presentation by emphasizing two observational
circumstances: i) It may be pivotal in which energy range relevant for coronal losses 
the observations are made. Observations that exclusively record the harder part of 
soft X-rays selectively favor detections of large flares and
suppress the relevance of low-energy flares (due to the Feldman et al.
relation; see also discussion in \citealt{porter95}). 
As is to be expected from the flare-heating hypothesis, the quiescent 
emission is comparatively soft and is therefore also underrepresented in hard 
observations. We  have marginally found this effect in our MECS observations.
One may wonder whether a similar effect exists for non-thermal hard X-rays
often used for solar flare energy statistics. If they are generated overproportionally  in
larger flares (as suggested by the ``Big Flare Syndrome'', \citealt{kahler82}, 
but also by recent observations
finding that microflares are radio-poor, i.e., relatively weak in the production of
accelerated particles, \citealt{krucker00}) then the statistical distributions may be biased toward too low $\alpha$.
We have selected  the energy range in which the dominant fraction of the coronal flare energy
is radiated. Also, the efficiency (ratio between observed count rate and incident flux) of the 
DS and the LECS detectors shows only a  weak temperature dependence, i.e., the
observations are equally sensitive to plasma over a wide range of relevant temperatures.
ii) The power-law distribution may depend on the flare energy range
considered. There are indications in solar observations to this effect, and we can safely state
that the power laws found here cannot be extrapolated to arbitrary energies: There must be a low-energy
break (possibly changing to a shallower distribution) in order to confine the total radiated
power, and there must be a high-energy limit, corresponding to the largest
physically possible flares.

\acknowledgments
The authors thank the referee, Alexander Brown, for thoughtful and constructive comments
on the original manuscript. We thank the {\it EUVE} and {\it BeppoSAX} staff for 
their generous support of our
observing campaign. Th PSI group (MG, MA) acknowledges partial financial support by 
 the Swiss National Science Foundation (grants 2100-049343 and  2000-058827).
JJD and VLK acknowledge support from NASA grants and the Chandra X-Ray
Center during the course of this research.

\end{document}